\documentclass[useAMS,usenatbib]{mn2e}

\usepackage{amssymb,natbib,graphicx,subfigure,color}

\def\LCDM{$\Lambda\mbox{CDM}$}

\def\AREPO{{\small AREPO}}

\def\FOF{{\small FOF}}
\def\SUBFIND{{\small SUBFIND}}

\def\SFR{{\rm SFR}}

\newcommand{\be}{\begin{equation}}
\newcommand{\ee}{\end{equation}}
\newcommand{\bea}{\begin{eqnarray}}
\newcommand{\eea}{\end{eqnarray}}

\newcommand{\Fig}[1]{Fig.~\ref{f:#1}} 
\newcommand{\Figs}[2]{Figs.~\ref{f:#1} and \ref{f:#2}} 

\newcommand{\ifm}[1]{\relax\ifmmode#1\else$\mathsurround=0pt #1$\fi}

\newcommand{\sggg}[1]{\textcolor{green}{[]}}

\def\dex{{\rm\thinspace dex}}
\def\log{{\rm\thinspace log}}

\def\micron{{\rm\thinspace \mu m}}
\def\pc{{\rm\thinspace pc}}
\def\kpc{{\rm\thinspace kpc}}
\def\Mpc{{\rm\thinspace Mpc}}
\def\Mpccb{\,{\rm Mpc}^{3}}

\def\kms{{\rm\thinspace km\thinspace s}^{-1}}

\def\pMpccb{\,{\rm Mpc}^{-3}}

\def\erg{{\rm\thinspace erg}}

\def\Msun{\hbox{$\rm\thinspace M_{\odot}$}}
       
\def\yr{{\rm\thinspace yr}}

\def\Gyr{{\rm\thinspace Gyr}}
\def\Msunyr{\Msun\yr^{-1}}
\def\Msunpc2{{\Msun\pc}^{-2}}
\def\Msunyrkpc2{{\Msun\yr^{-1}\kpc}^{-2}}
\def\arcsec{{\rm\thinspace arcsec}}
\def\magarcsec2{{\rm\thinspace mag\thinspace arcsec}^{-2}}

\newcommand{\apj}{ApJ}
\newcommand{\apjl}{ApJL}
\newcommand{\apjs}{ApJS}
\newcommand{\aj}{AJ}
\newcommand{\mnras}{MNRAS}
\newcommand{\aap}{A\&A}

\newcommand{\nat}{Nature}

\title[The Illustris Simulation across cosmic time]{Introducing the Illustris Project: the evolution of galaxy populations across cosmic time}

\author[Genel, S. et al.]
{\parbox{20cm}{
Shy Genel$^{1}$\thanks{E-mail: sgenel@cfa.harvard.edu}, Mark Vogelsberger$^{2}$, Volker Springel$^{3,4}$, Debora Sijacki$^{5}$, Dylan Nelson$^{1}$,\\ Greg Snyder$^{6}$, Vicente Rodriguez-Gomez$^{1}$, Paul Torrey$^{1}$, and Lars Hernquist$^{1}$}\vspace{0.3cm}\\
$^{1}$Harvard-Smithsonian Center for Astrophysics, 60 Garden Street, Cambridge, MA 02138, USA\\
$^{2}$Department of Physics, Kavli Institute for Astrophysics and Space Research, Massachusetts Institute of Technology, Cambridge, MA 02139, USA\\
$^{3}$Heidelberg Institute for Theoretical Studies, Schloss-Wolfsbrunnenweg 35, 69118 Heidelberg, Germany\\
$^{4}$Zentrum f{\"u}r Astronomie der Universit{\"a}t Heidelberg, ARI, M{\"o}nchhofstr. 12-14, 69120 Heidelberg, Germany\\
$^{5}$Institute of Astronomy and Kavli Institute for Cosmology, Cambridge University, Madingley Road, Cambridge CB3 0HA, UK\\
$^{6}$Space Telescope Science Institute, 3700 San Martin Drive, Baltimore, MD 21218, USA}

\begin{document}

\maketitle

\label{firstpage}

\begin{abstract}
We present an overview of galaxy evolution across cosmic time in the Illustris Simulation. Illustris is an N-body/hydrodynamical simulation that evolves $2\times1820^3$ resolution elements in a $(106.5\Mpc)^3$ box from cosmological initial conditions down to $z=0$ using the \AREPO{ }moving-mesh code. The simulation uses a state-of-the-art set of physical models for galaxy formation that was tuned to reproduce the $z=0$ stellar mass function and the history of the cosmic star-formation rate density. We find that Illustris successfully reproduces a plethora of observations of galaxy populations at various redshifts, for which no tuning was performed, and provide predictions for future observations. In particular, we discuss {\bf (a)} the buildup of galactic mass, showing stellar mass functions and the relations between stellar mass and halo mass from $z=7$ to $z=0$, {\bf (b)} galaxy number density profiles around massive central galaxies out to $z=4$, {\bf (c)} the gas and total baryon content of both galaxies and their halos for different redshifts, and as a function of mass and radius, and {\bf (d)} the evolution of galaxy specific star-formation rates up to $z=8$. In addition, we {\bf (i)} present a qualitative analysis of galaxy morphologies from $z=5$ to $z=0$, for the stellar as well as the gaseous components, and their appearance in HST mock observations, {\bf (ii)} follow galaxies selected at $z=2$ to their $z=0$ descendants, and quantify their growth and merger histories, and {\bf (iii)} track massive $z=0$ galaxies to high redshift and study their joint evolution in star-formation activity and compactness. We conclude with a discussion of several disagreements with observations, and lay out possible directions for future research.
\end{abstract}


\section{Introduction}
\label{s:intro}
Even though we have witnessed immense progress in our observational knowledge of galaxies over the past three decades, galaxy formation theory is not yet fully predictive. This situation has its roots in the strong non-linearity of the dynamics and the vast range of scales and physical effects that couple together and play a crucial role in shaping galaxies. Obtaining theoretical predictions for the abundance, spatial distribution, internal properties, and evolution of galaxies requires numerical simulations that follow the main components of the Universe, namely dark energy, dark matter (DM), and normal (baryonic) matter from cosmological initial conditions. While the evolution of the dark components is usually assumed to be governed only by gravitational interactions, which can be simulated to high accuracy, the baryons are clearly subject to the operation of a multitude of physical processes, several of which are extremely complex and poorly understood.

A useful approach for tackling this complexity is to make a series of simplifications to various components of the physics, which necessarily introduce free parameters into the models and reduce their predictive power. Gravity is the basic force that drives structure formation in the Universe, and luckily it is also relatively simple to model \citep{BarnesJ_86a,SpringelV_05a}, hence it is included in any cosmological simulation. A dividing line between two common numerical techniques lies in the modeling of the hydrodynamics of cosmic gas. With the so-called semi-analytical technique (e.g.~\citealp{KauffmannG_93b,ColeS_94a,DeLuciaG_07a,BensonA_10a}), hydrodynamics is not followed explicitly, but is instead prescribed using crude phenomenological recipes, which renders this technique computationally inexpensive. On the other side of the dividing line stand hydrodynamical cosmological simulations that solve Euler's equations, where the baryons are modeled self-consistently with the dark components \citep{HernquistL_89b,KatzN_92a,CenR_92b}. These simulations have a significantly increased level of predictive power, and are the state of the art for cosmological simulations. Additional physical components and processes, such as chemistry, star-formation, radiation, magnetic fields, cosmic rays, and the high-energy physics around black holes affect the formation of galaxies in a significant way, so that they must be modeled in a coarse, parameterised form, which in the context of hydrodynamical simulations is often termed `sub-resolution'.

Given the significant difference between semi-analytical models and hydrodynamical simulations, they have been used for largely orthogonal purposes. Semi-analytical models, thanks to their higher flexibility in physics modeling and significantly lower computational cost, have targeted an ambitious goal: reproducing, in a statistical sense, all available observations of galaxies \citep{GuoQ_11a,SomervilleR_08a,SomervilleR_12a}. Progress is being made by tuning the free parameters of the models and testing the results against observations \citep{HenriquesB_13a}. However, hydrodynamical cosmological simulations, due to their limited flexibility and high computational cost with respect to semi-analytical models, have usually targeted more focused applications \citep{DaveR_08a,DeasonA_11a,OserL_11a,RahmatiA_13a}, sometimes even in `zoom-in' simulations of a handful of particular objects \citep{GenelS_11a,ZolotovA_12,MarinacciF_14a,PlanellesS_14a}. They are often employed to study the effects of various models on a particular issue, without explicitly aiming at reproducing a multitude of observations \citep{OppenheimerB_06a,SchayeJ_10a,SalesL_10a,ChoiJ_10a,ScannapiecoC_12a,PuchweinE_13a}. Only very recently, have exceptions to that rule appeared \citep{Brook_C12a,KannanR_14a,KhandaiN_14a,SchayeJ_14a}, with encouraging, even if still limited, success.

\begin{figure}
\centering
\includegraphics[width=0.475\textwidth]{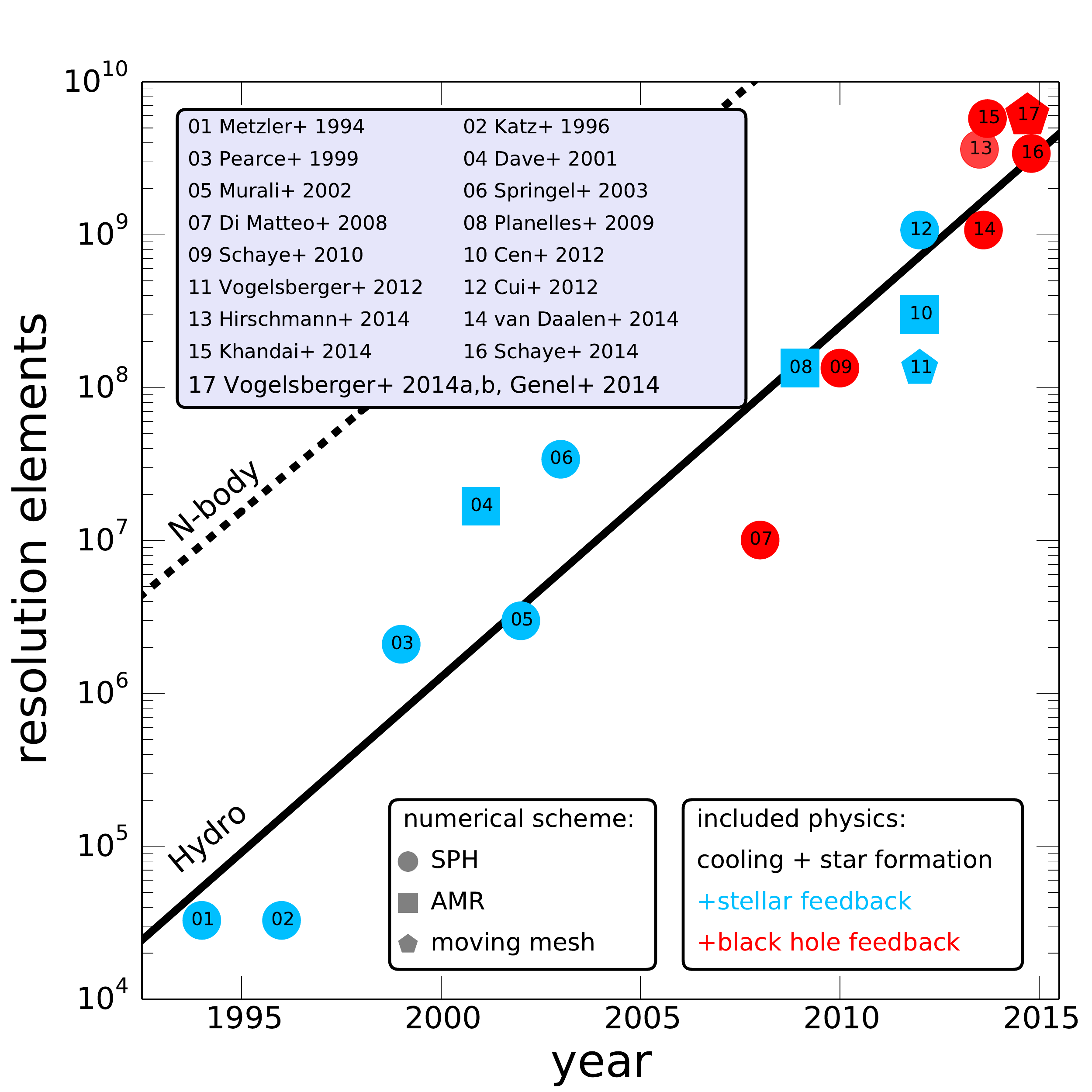}
\caption{Number of hydrodynamical resolution elements of uniform-box cosmological simulations that are evolved down to $z=0$, as a function of publication date. Different symbols indicate different hydrodynamical schemes (SPH, AMR, and moving mesh). Most simulations use the SPH technique, while only a handful have employed finite volume schemes. We only show simulations that use a spatially-adaptive hydrodynamical scheme (i.e.~we exclude fixed-grid simulations), and include at least cooling, star-formation and some form of star-formation feedback. Only recently have simulations begun to include AGN feedback as well (red). The solid black line shows an exponential best-fit to the growth of the number of resolution elements of hydrodynamical simulations, with a doubling time of $16.2$ months. For comparison, the black dashed line shows the growth of pure N-body simulations, with a doubling time of $16.5$ months \citep{SpringelV_05a}.}
\vspace{0.3cm}
\label{f:simulations_history}
\end{figure}

The frontier in cosmological hydrodynamical simulations since they came to existence about two decades ago has been pushed forward in three distinct directions. First, the size of the simulations, represented typically by the number of resolution elements, has been constantly increasing with more available computer power. This allows the combination of numerical resolution (required both for resolving fine details and for the robustness of numerical models) and cosmological volume (required for making statistically-meaningful comparisons with large observed data sets) to be ever improved \citep{PearceF_99a,DaveR_01a,SpringelV_03b,PlanellesS_09a,CuiW_12a,KhandaiN_14a,DuboisY_14a,SchayeJ_14a}. Second, the scope, complexity, and numerical robustness of sub-resolution models have been extended and improved upon, in order to provide a more complete and accurate description of the physical processes that shape galaxies, such as star-formation and feedback \citep{SpringelV_03a,OppenheimerB_06a,HopkinsP_11a}, black-hole growth and feedback \citep{SpringelV_05d,SijackiD_06a,BoothC_09a,DuboisY_12b}, magnetic fields \citep{DolagK_06a,PakmorR_14a}, and cosmic rays \citep{JubelgasM_08a,VazzaF_14a}. Finally, numerical techniques have been improved to provide more accurate solutions of the underlying equations of gravity and hydrodynamics \citep{BryanG_99a,SpringelV_02a,TeyssierR_02a,SpringelV_10a,ReadJ_12a,HopkinsP_13a}. These trends are demonstrated in \Fig{simulations_history}, which presents a historical summary of (non-`zoom-in') cosmological hydrodynamical simulations that were evolved down to $z=0$. The growth in the number of resolution elements can be reasonably well fit by an exponential with a doubling time of $16.2^{+4.1}_{-2.7}$ months (solid curve). Various numerical approaches are represented by different symbols and colours, which highlights that only recently have simulations begun to routinely incorporate AGN feedback and employ finite volume schemes.

In this work and two companion papers \citep{VogelsbergerM_14a,VogelsbergerM_14b} we present the Illustris Project, which represents an advance compared to previous work by combining the state-of-the-art in all three aforementioned areas -- number of resolution elements, scope of physical description, and numerical technique. The model used to perform the simulations has been tuned specifically so that the global efficiency of star-formation in the Universe roughly matches observations. While in \citet{VogelsbergerM_14a} and \citet{VogelsbergerM_14b} we focus on the $z=0$ simulated universe, here we study galaxy formation and evolution in the simulation from $z=0$ up to $z=8$, and compare the simulation results to observations. We also put particular emphasis on massive galaxies, which could not be probed in our earlier work that used a significantly smaller volume \citep{VogelsbergerM_13a,TorreyP_14a}.

This paper is structured as follows. In Section \ref{s:simulations} we present the specifications of the simulations. In Section \ref{s:cosmic_variance} we estimate the effect of cosmic variance on the basic characteristics of the galaxy populations in our simulation box. In Section \ref{s:stellar_mass} we discuss the buildup of the stellar mass in the simulation across cosmic time, in Section \ref{s:gas} we examine the baryon and gas contents of halos and galaxies, and Section \ref{s:sSFR} addresses the evolution of star-formation rates of galaxies. In Section \ref{s:morphologies} we present a qualitative account of basic galactic morphological properties. In Section \ref{s:mergertrees} we use the simulation to connect several $z=2$ galaxy populations to their $z=0$ descendants. In Section \ref{s:summary} we summarise our results and conclude with a discussion of directions for future research.

\section{The Simulations}
\label{s:simulations}
The Illustris project consists of a set of cosmological hydrodynamical simulations of a periodic box $106.5\Mpc$ on a side that is evolved to $z=0$ with various resolution levels and implemented physics. All simulations assume a standard \LCDM{ }cosmology with $\Omega_\Lambda=0.7274$, $\Omega_m=0.2726$, $\Omega_b=0.0456$, $\sigma_8=0.809$, $n_s=0.963$, and $H_0=70.4\kms\Mpc^{-1}$. The largest and most complete calculation in terms of implemented physical processes, Illustris-1, also referred to simply as the Illustris Simulation, follows $1820^3$ DM particles and approximately $1820^3$ baryonic resolution elements. It includes a comprehensive set of physics models for galaxy formation. In this simulation, the mass of a DM particle is $m_{\rm DM}=6.26\times10^6\Msun$ and the typical mass of a baryonic resolution element is $\overline{m}_b=1.26\times10^6\Msun$. The gravitational softening for DM particles is $1.4\kpc$ constant in comoving units, while for collisionless baryonic particles it is equal to that of the DM at $z\geq1$ and later on it is fixed to $0.7\kpc$, constant in physical units. For gas, the gravitational softening is tied to the cell size, with an imposed minimum equal to the value used for the collisionless baryonic particles. Two lower-resolution versions of the same simulation, using $2\times910^3$ and $2\times455^3$ resolution elements, Illustris-2 and Illustris-3 respectively, are also included in the set, mainly for resolution study purposes. In addition, the same box is simulated with less complete physics implementations, namely as a set of DM-only runs (Illustris-Dark), and a set of non-radiative (adiabatic) runs. All hydrodynamical runs also include passive Monte-Carlo tracer particles that allow the gas flow to be followed in a Lagrangian way. During the course of each run, $136$ snapshots were generated. The $61$ snapshots at $z>3$ are spaced with $\Delta\log( a)\approx0.02$, where $a$ is the cosmological scale factor, and the $75$ snapshots at $z<3$ are spaced with $\Delta a\approx0.01$. Each snapshot was post-processed by extended and optimized versions of the \FOF{ }and \SUBFIND{ }algorithms \citep{DavisM_85a,SpringelV_01,DolagK_09a}, to identify halos and subhalos and their properties.

The Illustris simulations were evolved with the moving-mesh code \AREPO{ }\citep{SpringelV_10a}, which solves the Euler equations with Godunov's scheme on a quasi-Lagrangian moving Voronoi mesh. We have shown this method to be advantageous in terms of accuracy compared to the traditional methods of smoothed particle hydrodynamics (SPH) and adaptive mesh refinement (AMR) \citep{VogelsbergerM_12a,SijackiD_12a,KeresD_12a,TorreyP_12a,GenelS_13a,NelsonD_13a}. The full details of the physical models used in Illustris are described in \citet{VogelsbergerM_13a}, and the Monte-Carlo tracer particle technique is detailed in \citet{GenelS_13a}. 

The physics modeling includes, on top of gravity, hydrodynamics, and a uniform ionizing background \citep{KatzN_96b,FaucherGiguereC_09a} in an expanding universe, a set of sub-resolution models for processes that are important for galaxy formation. These include radiative cooling (primordial as well as by heavy elements) with self-shielding corrections, star-formation in high-density gas, stellar evolution with associated metal enrichment and mass return, star-formation feedback that uses $1.09\times10^{51}\erg$ per supernova explosion to drive kinetic galactic winds, black hole seeding, accretion and merging, and finally three distinct types of AGN feedback: quasar-mode, radio-mode, and a radiative mode. The $15$ or so free parameters of the models, associated mostly with the various feedback processes, all have a physical meaning, and can be assigned numerical values based on underlying principles, but given our ignorance and uncertainties regarding the complicated physics of, e.g.~star-formation and black-hole accretion, there is freedom in their exact values. In practice, a subset of them was tuned to their particular values based on test simulations (of a much smaller volume, $35.5\Mpc$ on a side), which we only compared against less than a handful of basic observations, in particular the history of cosmic star-formation rate (SFR) density and the $z=0$ stellar mass function, ensuring that they were roughly reproduced. Full details on the physics models and the choice of parameters appear in \citet{VogelsbergerM_13a}.

\section{Simulation volume and cosmic variance}
\label{s:cosmic_variance}
The finite computational resources available for any simulation impose limits on the mass, spatial, and temporal scales that can be simulated and resolved. For periodic-box (non-`zoom-in') cosmological simulations, the main compromises that can be made to render the calculation tractable are the cosmic time that is followed and the number of resolution elements. The latter, in turn, constrains the combination of simulation box size and mass resolution. In choosing the parameters for Illustris, the available computational resources together with our goal of running the simulation to $z=0$ fixed the number of resolution elements at $2\times1820^3$. Further, the combination of this number of total resolution elements with our choice to resolve $L^*$ galaxies with tens of thousands of resolution elements each (which allows to reach convergence for those systems as well as to obtain a spatial resolution of $\kpc$ scales), is what determined the size of the Illustris simulation volume.

The resulting box side length of $106.5\Mpc$, is, as argued below, in a `transition region' between small boxes of sizes that deviate from linear growth at low redshift and suffer from strong cosmic variance, and large boxes that remain `homogeneous' down to $z=0$ and contain statistical samples even of cluster-scale DM halos (see also \citealp{BaglaJ_05b}). As shown in \citet{VogelsbergerM_14b}, the most massive halo in the Illustris box has $M_{200c}\approx2.3\times10^{14}\Msun$, which means that the volume is not large enough to contain the most massive halos, the ones that host galaxy clusters such as the Virgo cluster. However, the number of `cluster-scale' halos with $M_{200c}>10^{14}\Msun$ is $10$, and the number of `Milky-way-like' halos with $10^{12}\Msun<M_{200c}<2\times10^{12}\Msun$ (roughly corresponding to the \citet{PifflT_14a} estimate) is $733$. The fact that Illustris contains hundreds of $L^*$ galaxies, each resolved with tens of thousands of resolution elements, as well as several small clusters, each resolved tens of millions of resolution elements, all in one self-consistent volume, is a direct result of the large number of resolution elements evolved in the simulation down to $z=0$.

The initial conditions were selected manually from a set of $100$ trials that represent different random realizations of the Illustris cosmology in a $106.5\Mpc$ box, each evolved using a low-resolution DM-only simulation to $z=0$. The criteria for choosing one out of those hundred were that the DM halo mass function should not deviate strongly from the theoretical mass function (e.g.~\citealp{JenkinsA_01a}), and that the box is relatively spatially homogeneous upon a visual inspection. In this crude way, we attempted to run Illustris as a `representative' volume of $1.2\times10^6\Mpccb$ (see the Illustris halo mass function in \citet{VogelsbergerM_14b}). We will now investigate to what degree cosmic variance on the scale of $106.5\Mpc$ is expected to affect some basic baryonic characteristics of the Illustris volume. Specifically, we will estimate the expected cosmic variance of the stellar mass functions at several redshifts, and of the history of the cosmic SFR density $\rho_{\rm SFR}$.

We determine the cosmic variance sensitivity of those quantities utilizing an estimate of the cosmic variance of the DM halo mass function that we extract from the Millennium simulation\footnote{We did not use the Millennium simulation as is, but first converted it to the WMAP-7 cosmology used in Illustris following \citet{AnguloR_10a} and \citet{AnguloR_12a}.} \citep{SpringelV_05a}. We first select from the Millennium simulation $6^3$ mutually-exclusive cubic sub-volumes of $106.5\Mpc$ on a side. We also define mutually-exclusive bins of halo mass, and for each of the sub-volumes, we use those bins to quantify the halo mass function. Then we record the difference (in logarithmic space) between the $16$th, $50$th, and $84$th percentiles of the halo mass function among the various sub-volumes. This procedure is repeated for different redshifts. In the second step, we calculate the contribution of each of the mutually-exclusive halo mass bins to the stellar mass function and to the history of cosmic SFR density from the full Illustris volume. In addition, we record a `$1\sigma$-higher' (`$1\sigma$-lower') contribution from each bin, by adding, in logarithmic space, the corresponding $84$th percentile-to-median ($16$th percentile-to-median) difference that we obtained in the first step to the actual contribution of that bin in Illustris\footnote{For halo mass bins that are resolved in Illustris but are below the Millennium simulation resolution, we use the cosmic variance estimate at the lowest mass resolved by the Millennium simulation, $\approx10^{10.5}\Msun$. This gives only a small over-estimate of the cosmic variance for those bins, since at such low masses the cosmic variance is almost independent of mass.}. Finally, we construct stellar mass functions and the history of cosmic SFR density from Illustris by summing up the contributions to these quantities coming from the various halo mass bins. This is done both for the actual values measured from Illustris (which gives a result identical to simply taking all halos without summing them up in bins first) as well as for the `$1\sigma$-higher' and `$1\sigma$-lower' contributions. Thereby, we obtain an estimate for each of those baryonic quantities had they been built from DM halo mass functions that deviate from the Illustris halo mass functions systematically by $\approx1\sigma$.

However, for these systematically-offset quantities to exactly represent $1\sigma$ variations around the actual values, several assumptions need to be made: {\bf(a)} the distribution among the sub-volumes is lognormal, {\bf(b)} the halo mass bins are independent, {\bf(c)} the Illustris halo mass function is exactly the median one, and {\bf(d)} the baryonic quantities in question scale linearly with the halo mass. Since it is clear that these assumptions do not strictly hold, but are only approximations (see, e.g.~\citealp{XieL_14a}), we first verify that this approach provides reliable estimates. To do that, we repeat the procedure described above, but instead divide the Millennium simulation into $18^3$ sub-volumes in order to estimate the cosmic variance on a $35.5\Mpc$ scale\footnote{We also verified that the `typical' deviations as a function of mass and redshift obtained in this way are very similar to the ones obtained by dividing the Illustris volume itself into $3^3$ sub-volumes of $35.5\Mpc$.}.

\begin{figure*}
\centering
\subfigure[Cosmic SFR density, $35.5\Mpc$ cosmic variance]{
          \label{f:cosmic_variance_SFRH_25Mpc}
          \includegraphics[width=0.49\textwidth]{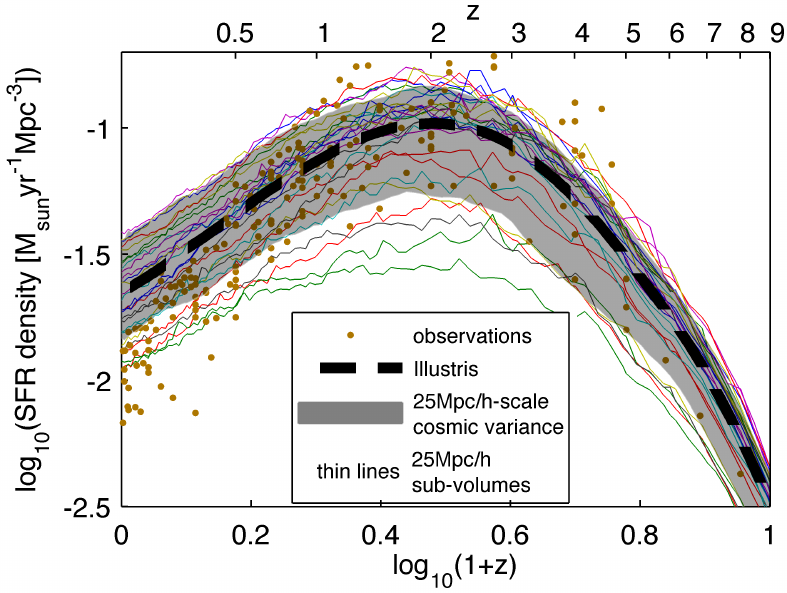}}
\subfigure[$z=0$ stellar mass function, $35.5\Mpc$ cosmic variance]{
          \label{f:cosmic_variance_SMF_25Mpc}
          \includegraphics[width=0.49\textwidth]{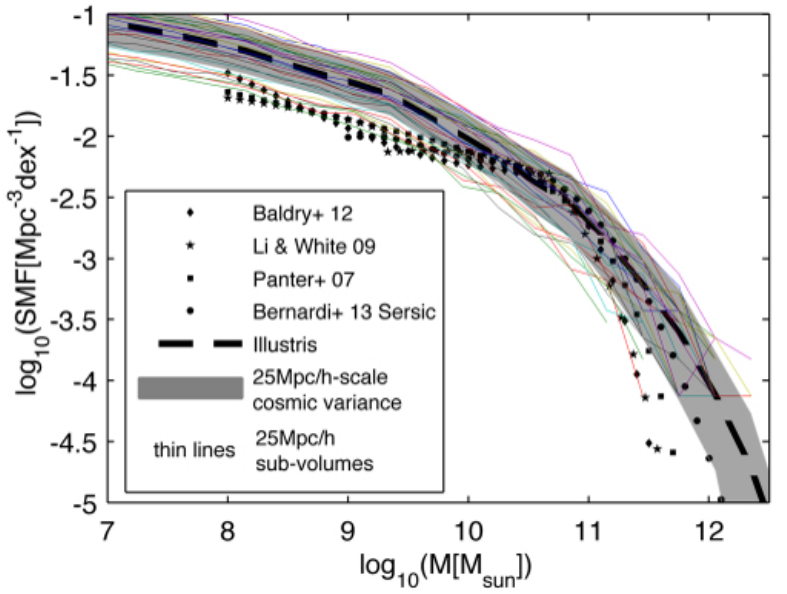}}
\subfigure[Cosmic SFR density, $106.5\Mpc$ cosmic variance]{
          \label{f:cosmic_variance_SFRH_75Mpc}
          \includegraphics[width=0.49\textwidth]{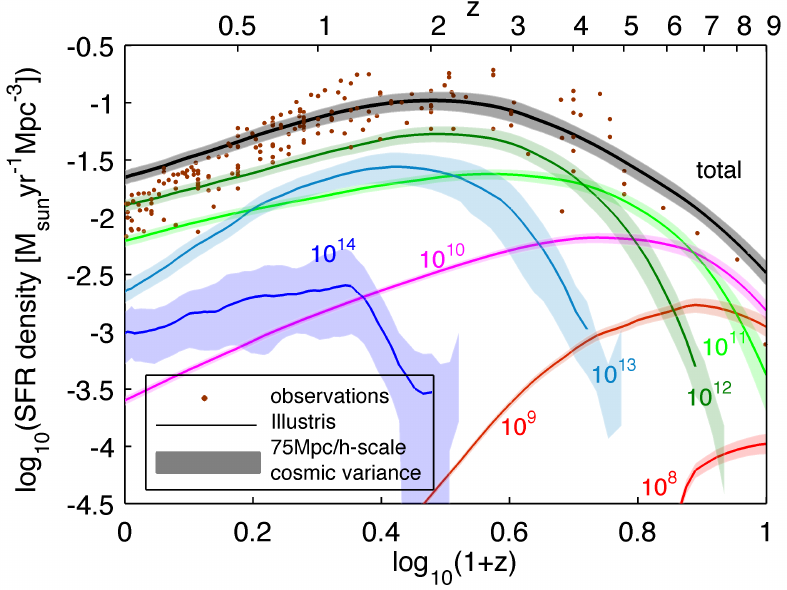}}
\subfigure[Stellar mass functions, $106.5\Mpc$ cosmic variance]{
          \label{f:cosmic_variance_SMF_75Mpc}
          \includegraphics[width=0.49\textwidth]{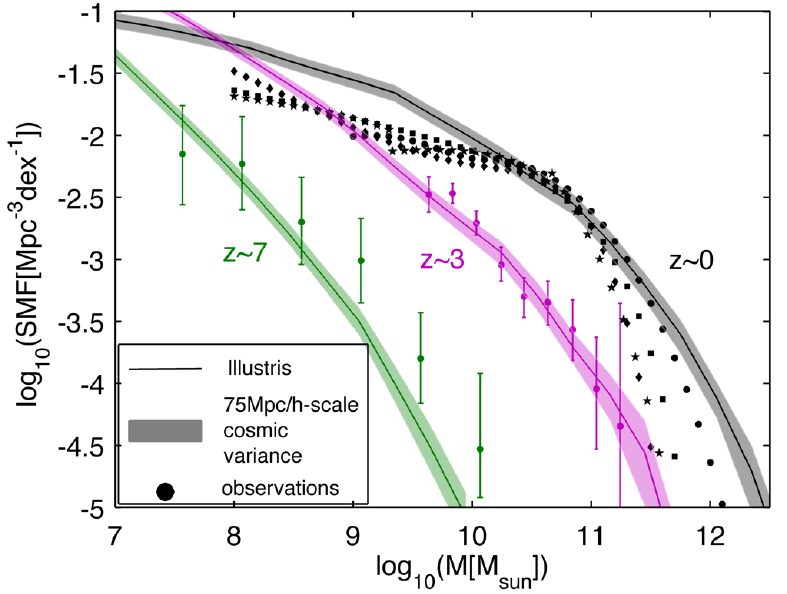}}
\caption{The effect of cosmic variance on the history of cosmic SFR density (left panels) and on the stellar mass function (right panels). The top panels present a study of our procedure for determining the cosmic variance on the scale of $35.5\Mpc$ (shaded regions) around the total result from Illustris (dashed curve). This estimate for the cosmic variance is in agreement with the results from $3^3$ mutually-exclusive $(35.5\Mpc)^3$ sub-volumes of the full Illustris volume (thin curves). The bottom panels show the estimated cosmic variance effects on the full box scale, i.e.~$106.5\Mpc$, which are found to be much smaller compared to those on the $35.5\Mpc$ scale. These are mostly also smaller than the observational uncertainty. Hence, the Illustris volume is large enough so that our results are not significantly affected by cosmic variance.}
\vspace{0.3cm}
\label{f:cosmic_variance}
\end{figure*}

In \Fig{cosmic_variance_SFRH_25Mpc} we present the history of cosmic SFR density $\rho_{\rm SFR}$ for Illustris (dashed thick curve) with its $35.5\Mpc$-scale cosmic variance estimate (shaded region). We compare this estimate with the actual SFR density histories for $27$ mutually-exclusive sub-volumes of the Illustris volume (thin curves). It is apparent that roughly $68\%$ of the sub-volumes are inside the cosmic variance estimate obtained with our procedure, as expected. A similar exercise is shown in \Fig{cosmic_variance_SMF_25Mpc} for the $z=0$ stellar mass function, with the same conclusion, namely that our procedure produces reliable estimates of the cosmic variance.

Before discussing the cosmic variance on $106.5\Mpc$ scales, it is worth commenting here with respect to the actual magnitude of cosmic variance on the $35.5\Mpc$ scale. Both for $\rho_{\rm SFR}$ and for the $z=0$ stellar mass function it is quite large, both in absolute terms (root-mean-square deviations of factor $\approx2$), and more importantly, compared to the observational uncertainties. Since the parameters of our galaxy formation physics models were tuned to reproduce those two quantities using a $35.5\Mpc$ box \citep{VogelsbergerM_13a}, it should be expected that these parameter choices suffer from the effects of cosmic variance. Indeed, the history of cosmic SFR density and stellar mass function as tuned for on the $35.5\Mpc$ box from \citet{VogelsbergerM_13a} are not exactly reproduced in the Illustris volume. In fact, the particular $35.5\Mpc$ box we used can now be judged as somewhat `unlucky' in the sense that both quantities are approximately $0.1\dex$ lower in that box compared to Illustris, such that Illustris moved $\approx0.1\dex$ away from the observational data in the upwards direction.

In \Figs{cosmic_variance_SFRH_75Mpc}{cosmic_variance_SMF_75Mpc} we present $\rho_{\rm SFR}$, and the stellar mass functions at redshifts $z=0,3,7$, respectively, together with the estimates for cosmic variance on the scale of the full simulation box. It is clear by comparison to the upper panels that the cosmic variance is dramatically reduced on the $106.5\Mpc$ scale compared with the $35.5\Mpc$ scale. In particular, the cosmic variance is smaller than the current observational uncertainties\footnote{The cosmic variance shaded regions for the mass functions may upon first impression appear to have a constant `width', which may be misleading. In fact, a careful inspection will clearly show that in the vertical direction the cosmic variance increases with stellar mass, as expected for rarer systems.}. This justifies our statement above that a cosmological box with a side length of $106.5\Mpc$ has a large enough volume so that it does not suffer from detrimental cosmic variance effects. We expect these two quantities to be potentially most significantly affected by cosmic variance, certainly more than scaling relations between internal galaxy properties, which are examined throughout the rest of this paper. Therefore, we conclude that the results we present hereafter are likely only mildly sensitive to cosmic variance, and hence we will not consider such effects throughout the rest of this work.

\section{Buildup of the stellar mass}
\label{s:stellar_mass}
\subsection{Stellar mass functions and the relations between stellar mass and halo mass}
\label{s:SMF_MSMH}
In this section we examine how the stellar mass builds up across cosmic time in the Illustris simulation. We begin by examining the history of cosmic SFR density shown in \Fig{cosmic_variance_SFRH_75Mpc}, where it is also broken up into contributions from different halo mass bins. By tuning the free parameters of the various feedback models \citep{VogelsbergerM_13a}, we were able to obtain a good match at $z>1$. However, the decrease of $\rho_{\rm SFR}$ at $z\lesssim1$ is not as rapid in Illustris ($\propto(1+z)^{\approx1.8}$) as observed ($\propto(1+z)^{\approx2.8}$). As a result, the $z=0$ cosmic SFR density in Illustris is $0.15-0.45\dex$ higher than the various observational estimates. \Fig{cosmic_variance_SFRH_75Mpc} shows that the dominant contributors to $\rho_{\rm SFR}$ are halos of $M\approx10^{12}\Msun$ at all $z\lesssim4$, and in particular at $z\lesssim0.5$, with an additional contribution, smaller by approximately a factor of two, from halos of $M\approx10^{11}\Msun$ (for a breakdown by stellar mass see \citealp{VogelsbergerM_14b}). This is in qualitative agreement with predictions of empirical models \citep{ConroyC_09a,BehrooziP_13d,BetherminM_13a}. Future improvements to the modeling will require suppression of star-formation at the mass scale of $M\approx10^{11-12}\Msun$ to obtain a better quantitative agreement with the $z\lesssim1$ observations.

\begin{figure*}
\centering
\includegraphics[width=1.0\textwidth]{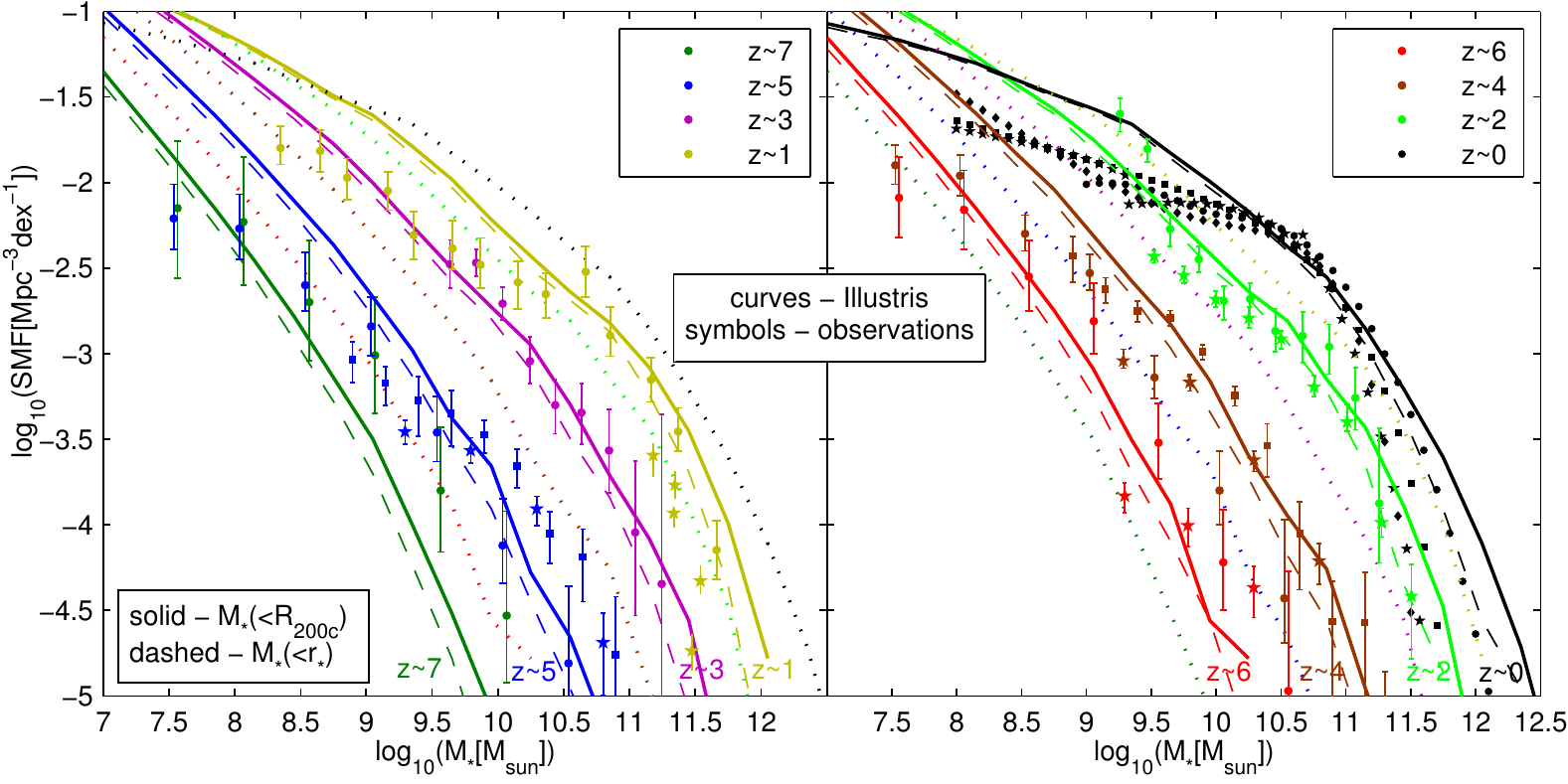}
\caption{The Illustris stellar mass function across cosmic time (curves) compared to observations (symbols, see below). Each panel presents a different set of redshifts to facilitate readability, but the solid curves from each panel are repeated as dotted curves in the other. We find overall agreement between Illustris and the observations, which is encouraging given that the free parameters of the feedback processes have not been tuned to match the $z>0$ mass functions. See discussion of remaining tensions with observations in the text. Two estimates for the galactic stellar mass are shown: the full stellar mass included in each \SUBFIND{ }halo (solid curves), and the stellar mass enclosed within the fiducial galactic radius $r_{\star}$ (dashed curves). The observational points are only a subset of the compilation presented in \citet{TorreyP_14a}, in order to keep the plots readable, however they fairly represent the full compilation both in normalization and scatter. The various observational points are adopted for $z\sim0$ from \citet{PanterB_07a} ({\it squares}), \citet{LiC_09a} ({\it stars}), \citet{BaldryI_12a} ({\it diamonds}), and \citet{BernardiM_13a} ({\it circles}), for $z\sim1$ from \citet{IlbertO_10a} ({\it stars}) and \citet{MortlockA_11a} ({\it circles}), for $z\sim2-3$ from \citet{KajisawaM_09a} ({\it stars}) and \citet{SantiniP_12a} ({\it circles}), for $z\sim4-5$ from \citet{LeeK_12a} ({\it squares}), for $z\sim4-6$ from \citet{StarkD_09a} ({\it stars}), and for $z\sim4-7$ from \citet{GonzalezV_11a} ({\it circles}).}
\vspace{0.3cm}
\label{f:SMF}
\end{figure*}

In \Fig{SMF} we present the stellar mass functions obtained in Illustris at various redshifts from $z=7$ to $z=0$, and compare them to observations. For ease of readability, we present odd redshifts ($z=7,5,3,1$) in the left panel, and even redshifts ($z=6,4,2,0$) in the right panel. We find overall excellent agreement between Illustris and the observations, in particular given that the free parameters that control the feedback processes in Illustris have been tuned to only approximately match the $z=0$ stellar mass function. However, we do find several tensions with respect to the observations that are worth noting: {\bf(a)} At the lowest redshifts the abundance at the massive end ($M_*\gtrsim10^{11}\Msun$) is somewhat overestimated (see discussion below and in Section \ref{s:z0MassiveEnd}). {\bf(b)} At the highest redshifts, $z\gtrsim6$, the abundance at the corresponding massive end ($M_*\gtrsim10^{9}\Msun$) is underestimated, however only by $\approx1\sigma$. {\bf(c)} At low redshift, $z\lesssim1$, the abundance below the knee ($M_*\lesssim10^{10}\Msun$) is overestimated by $\approx0.5\dex$ (see also \citealp{VogelsbergerM_14b}). {\bf(d)} At high redshift, $z\geq4$, the slope at the low-mass end ($M_*\lesssim10^{9}\Msun$) steepens with increasing redshift faster than observed. It is worth noting that the measurements for these low masses at high redshifts are taken from \citet{GonzalezV_11a}, whose mass function for the more massive galaxies ($M_*\gtrsim10^{9}\Msun$) is systematically lower than other measurements at $z\sim4$ (e.g.~\citet{StarkD_09a} that is shown here). For this reason, and also more generally, the observations at high redshift should be taken with caution with respect to systematics.

\begin{figure}
\centering
\includegraphics[width=0.475\textwidth]{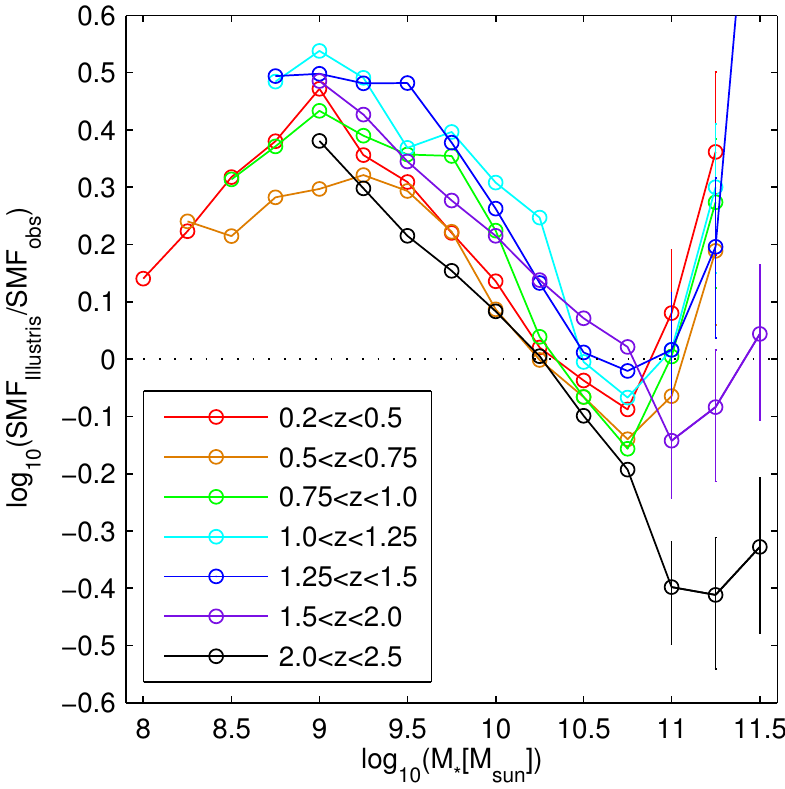}
\caption{The ratio between stellar mass functions in Illustris for a range of redshifts between $z=0.2$ and $z=2.5$, and the corresponding observed mass functions given by \citet{TomczakA_14a}. We find that the deviations of the simulated mass functions from the observed ones have a characteristic 'S' shape, which is generally preserved across this large redshift range. The error bars for the most massive bins represent the quoted error on the observational results. They are only shown for the most massive bins for visual clarity. For less massive bins, the error bars are similar to the ones shown for the $10^{11}\Msun$ bin.}
\vspace{0.3cm}
\label{f:SMF_ratio}
\end{figure}

Next, in \Fig{SMF_ratio} we focus our attention in more detail at the differences between the Illustris mass functions and observed ones at $z<2.5$, by showing the ratio between the two. We adopt the redshift bins reported in \citet{TomczakA_14a}, and obtain the corresponding mass functions from Illustris by averaging over the relevant snapshots. We find that the ratio between the simulated and observed mass functions have a rather well-defined 'S' shape, whose normalization depends on redshift. The best agreement occurs around the knee of the mass function, at $M_*\sim10^{10.5}\Msun$, where the Illustris mass function is equal to or lower than observed by $\approx0.1\dex$. For $M_*\gtrsim10^{11}\Msun$, the redshift dependence of the variations is approximately monotonic, with the difference between Illustris and the observations increasing at progressively lower redshifts. An extended discussion of stellar masses in the most massive halos at low redshift appears later in this section and in Section \ref{s:z0MassiveEnd}, but here we focus on the differences at the low-mass end. For $10^{9}\Msun\lesssim M_*\lesssim10^{10.5}\Msun$, Illustris overshoots the observed mass functions. This is increasingly the case at lower masses, which represents a steeper slope in Illustris than observed. However, in this mass range the redshift dependence is non-monotonic, with the largest differences found at $z\approx1$. At both lower and higher redshifts, the deviations show parallel sequences to that at $z\approx1$, offset towards lower values by $\approx0.25\dex$. This is a manifestation of the fact that the difference in slope between Illustris and observations is constant with redshift, even though both show a steepening towards higher redshifts. Both these aspects of our model are discussed also in \citet{TorreyP_14a}, based on a smaller cosmological volume. Finally, for $M_*\lesssim10^{9}\Msun$, the observed mass functions show a 'reverse knee' and steepen, a feature that is not reproduced in Illustris. This causes the difference between the two to start decreasing again towards the lowest masses. The various features in \Fig{SMF_ratio} are likely to represent different physical processes related either to the star-formation law or to the efficiency of galactic winds, or both. Flattening the 'S' shape to a constant (at zero difference) may be achieved with additional degrees of freedom in the models, which will be the topic of future research. It is encouraging that the shape of the deviations is largely similar at different redshifts in this wide redshift range.

\begin{figure}
\centering
\includegraphics[width=0.475\textwidth]{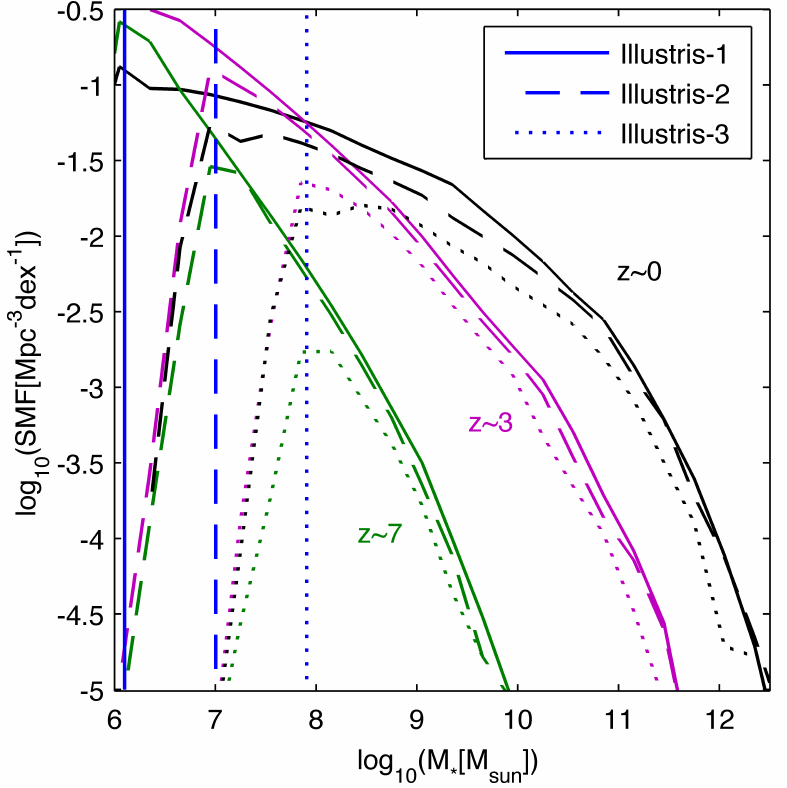}
\caption{Resolution study of the Illustris stellar mass function, which is shown for three different redshifts using the three resolution levels of the Illustris simulation suite. Three corresponding vertical lines denote the mean mass of a single baryonic resolution element in each of these simulations, $\overline{m}_b$. It can be seen that a consistent level of convergence holds all the way down to $\gtrsim\overline{m}_b$, and only at that scale the mass function is abruptly cut.}
\vspace{0.3cm}
\label{f:SMF_convergence}
\end{figure}

We examine the convergence of the stellar mass function with numerical resolution in \Fig{SMF_convergence}, where the different resolution levels of the Illustris simulation suite are shown by different line styles, for three redshifts $z=0,3,7$. These results demonstrate good convergence of galaxy stellar masses between Illustris-2 and Illustris-1, except for low-mass galaxies at low redshift, where the higher-resolution simulation still produces larger numbers of galaxies, consistently with our results in \citet{TorreyP_14a}. The lowest resolution simulation, Illustris-3, is significantly less well converged even in the most-resolved, high-mass end. Importantly, \Fig{SMF_convergence} also demonstrates that the convergence only breaks down for galaxy stellar masses that are very close to the typical mass of a single stellar particle, $\approx\overline{m}_b$. For Illustris-1, where $\overline{m}_b=1.26\times10^6\Msun$, this means that the mass functions presented in \Fig{SMF}, i.e.~for $M_*>10^7\Msun$, are quite robust to numerical resolution effects.

Further physical insight into the evolution of galactic stellar masses can be extracted from the relation between stellar mass and halo mass. This relation can be probed by direct observations of individual systems, using techniques such as weak lensing \citep{MandelbaumR_06b}, cluster atmosphere modeling \citep{GonzalezA_13a} or satellite kinematics \citep{ConroyC_07a}, and it can also be inferred with indirect methods employing large galaxy populations, e.g.~based on galaxy abundance and clustering and a combination thereof \citep{TinkerJ_13a}. A common indirect technique to determine this relation is dubbed `abundance matching', where the stellar mass function is matched to a theoretical halo mass function under the assumption of similar rank-ordering, with the possibility of allowing for some scatter \citep{ConroyC_09a,MosterB_09a,GuoQ_10a,BehrooziP_10a,BehrooziP_13b}. While it is encouraging that various studies, using these completely independent techniques, agree with one another to within a factor of roughly two \citep{LeauthaudA_12a}, rather substantial differences persist, even between different studies utilizing the same technique, such as abundance matching.

\begin{figure*}
\centering
\subfigure[Stellar mass to halo mass relation]{
          \label{f:MsMhzallR200Illustris}
          \includegraphics[width=0.49\textwidth]{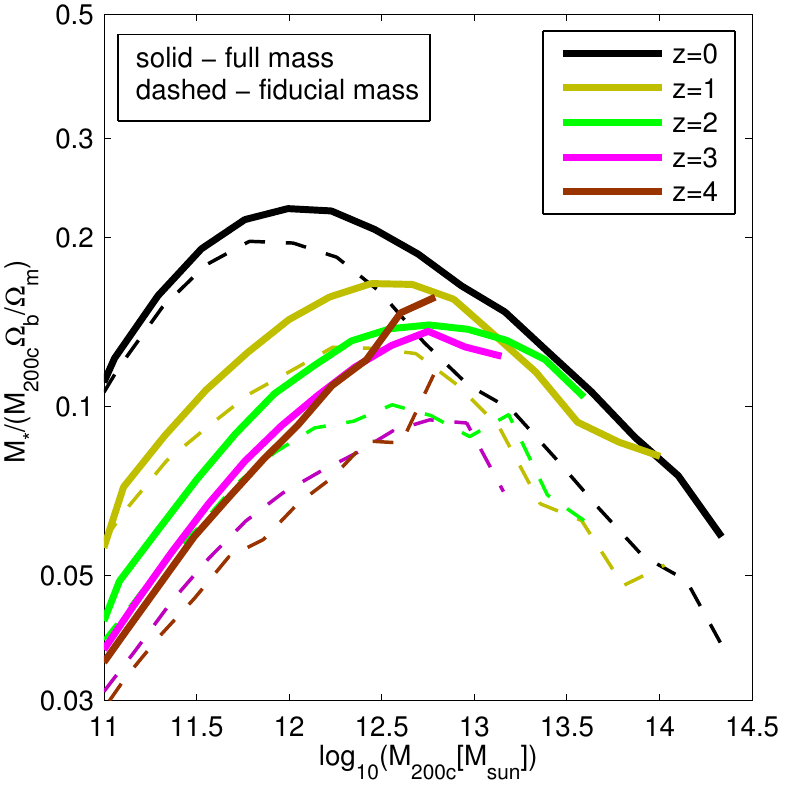}}
\subfigure[Effects of physics variation]{
          \label{f:MsMhzallR200L25comparison}
          \includegraphics[width=0.49\textwidth]{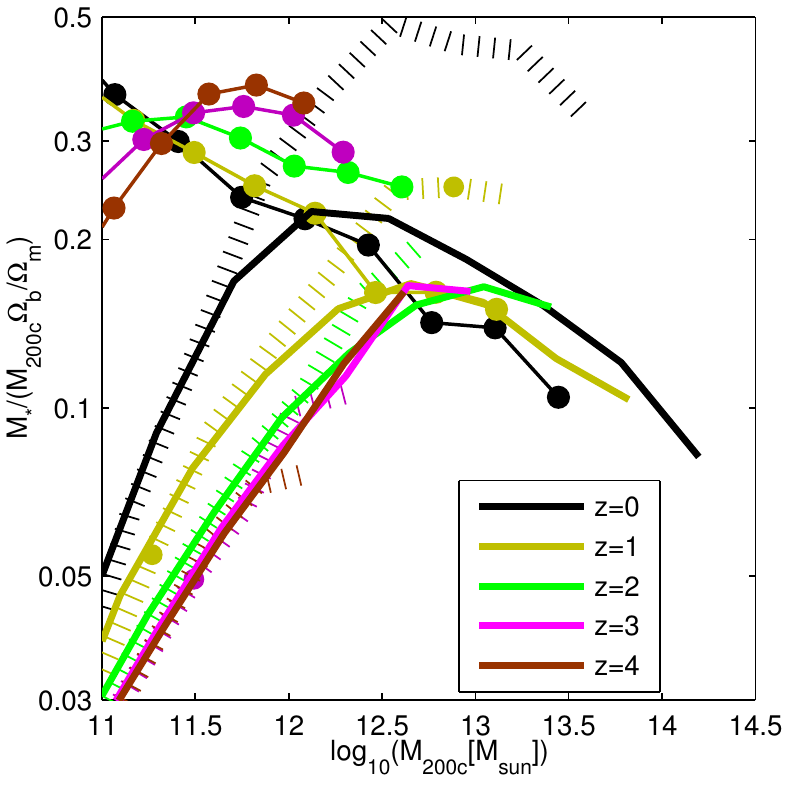}}
\caption{{\it Left panel:} The fraction of baryons that turned into stars as a function of halo mass, or the `baryon conversion efficiency', for central galaxies at redshifts $0\leq z\leq4$. Two curves are shown for each redshift, corresponding to the two types of curves in \Fig{SMF}: for the full stellar mass inside $R_{200c}$ excluding satellites (solid curves), and for the stellar mass within the fiducial galactic radius $r_{\star}$, again excluding satellites (dashed curves). The halo masses we use are the halo masses of individually-matched halos from Illustris-Dark, in order to make a fair comparison to abundance matching results, which use DM-only simulations for the halo mass (\citealp{SawalaT_13a}, Rodriguez-Gomez et al.~in prep.). {\it Right panel:} The same quantity is shown for several simulations that use a factor of $8$ lower mass resolution: the Illustris volume simulated with $2\times910^3$ resolution elements instead of $2\times1820^3$ (thick solid curves), the $35.5\Mpc$ box from \citet{VogelsbergerM_13a} with the same physical modeling as Illustris except the lack of radio-mode AGN feedback (dashed curves), and the same $35.5\Mpc$ box with the full physical modeling except the lack of galactic winds (solid thin curves with circles).}
\vspace{0.3cm}
\label{f:MsMhzallR200}
\end{figure*}

In \Fig{MsMhzallR200Illustris} we show the relation between stellar mass and halo mass in Illustris for central galaxies at various redshifts. We find these results to be in a good qualitative agreement at $z\lesssim3$ with the relations found by \citet{BehrooziP_13b} in that the peak of the baryon conversion efficiency moves to higher masses at higher redshifts, the efficiency at masses larger than the peak is independent of redshift, and the efficiency at masses smaller than the peak is decreasing with redshift. However, in quantitative detail the results disagree, as the dependence of the peak location on redshift is significantly stronger in our simulation. Also, the aforementioned trends with redshift are reversed according to \citet{BehrooziP_13b} at $z\gtrsim4$, a behaviour we do not find in our simulation, which instead seems to converge to a non-evolving relation at high redshifts.

The relations derived by \citet{MosterB_13a} have somewhat different qualitative and quantitative trends from those of \citet{BehrooziP_13b}. There, the shift of the peak towards higher masses at higher redshifts is much more pronounced, and is in better agreement with Illustris (similar results were found by the abundance matching study of \citealp{ConroyC_09a}). The baryon conversion efficiency at masses beyond the peak is rather constant, in agreement with both \citet{BehrooziP_13b} and Illustris. However, the behaviour at masses below the peak is rather different; instead of a constant slope and a decreasing normalization towards higher redshifts as in \citet{BehrooziP_13b} and in Illustris, \citet{MosterB_13a} find shallower slopes at higher redshifts. To summarize the comparison to results derived by abundance matching, we find that the evolution of the stellar mass to halo mass relation shares some features with the \citet{BehrooziP_13b} results and some with \citet{MosterB_13a}, and where both abundance matching results agree, the agreement is shared by Illustris as well. Therefore, we conclude that the stellar mass-halo mass relation obtained in Illustris is evolving in very good agreement with what is inferred from observations.

It is interesting to note that all the abundance matching studies we considered find a non-evolving (within the uncertainties) stellar mass-halo mass relation at masses beyond the peak of the `baryon conversion efficiency' curve (but see \citealp{LeauthaudA_12a}). This is a feature that Illustris reproduces in good agreement with the empirical models, without involving any tuning. This calls for a better understanding of the physical origins of this non-evolution. We make a first step in this direction by examining this relation in simulations that vary the physical models compared to our fiducial model. Such simulations, based on a $35.5\Mpc$ box, and with a mass resolution $\approx8$ times lower than in Illustris, were presented in \citet{VogelsbergerM_13a}. 

In \Fig{MsMhzallR200L25comparison} we first show that this lower mass resolution is well-enough converged in this respect by showing the same relation based on Illustris-2, a lower-resolution version of Illustris that was evolved with $2\times910^3$ resolution elements (thick solid curves). In the high-mass regime we are interested in, this simulation agrees very well with the higher-resolution Illustris-1. Next, we show the relation based on a simulation where no radio-mode AGN feedback was included but where otherwise the model is identical to the one used in Illustris (dashed curves). We find that the stellar mass-halo mass relation evolves strongly with redshift at all masses, including at the massive end. At high masses, these curves are similar to ones derived from a simulation with no feedback at all (not shown), as galactic winds have low mass-loading factors for high masses. This demonstrates that the constancy of the high-mass end of the `baryon conversion efficiency' in Illustris is not a natural result of cosmological accretion and radiative cooling, but requires some feedback. We then examine a simulation that does include radio-mode feedback but lacks galactic winds (thin solid curves with circles). We find the surprising result that the direction of the redshift evolution is reversed at the high-mass end, such that at low redshift the `baryon conversion efficiency' is lower than at high redshift. This can be interpreted as a result of the late operation of the radio-mode feedback, which kicks in only when the accretion rates drop. Compared to this simulation, the galactic winds that are included in the full fiducial model eject gas at high redshift that is later recycled at lower redshift, thereby canceling out the reverse dependence on redshift that is caused by the radio-mode feedback \citep{OppenheimerB_10a}. Thus, we find that it is only a combination of galactic winds and radio-mode AGN feedback conspiring together that results in a non-evolving `baryon conversion efficiency' at high masses.

\subsection{Stellar components in the most massive halos}
\label{s:z0MassiveEnd}
The exact definition of what constitutes the stellar mass of a simulated galaxy becomes relevant to the comparison with observations at low redshift, where the observational uncertainties are smaller. In \Fig{SMF} we show the stellar mass functions for two such definitions. The solid curves are based on the full stellar mass associated with the \SUBFIND{ }halo of each galaxy, which for central galaxies includes all the bound mass out to roughly the virial radius of their host halo, excluding satellite galaxies. The dashed curves are based on an assignment of the stellar mass for each galaxy as the stellar mass that is enclosed within twice the stellar half-mass radius of its \SUBFIND{ }halo, which is the radius that we use as a fiducial `galactic radius' and denote $r_{\star}$. Indeed, neither of these definitions corresponds exactly to observational criteria, but they may reasonably capture the typical variation that is expected due to various radial cuts (for a recent detailed discussion of the `edge' of simulated galaxies, see \citealp{StevensA_14a}). In fact, the `edge' of observed galaxies is not well-defined either, and in particular for massive galaxies the radius out to which the mass is integrated can strongly affect the measured stellar mass \citep{BernardiM_13a}. We find that for low-mass galaxies, there is negligible difference between the mass functions derived with these two definitions, while for massive galaxies, the resulting mass function can be shifted by up to $\approx0.2\dex$ in the horizontal direction, a difference that is sub-dominant compared with observational uncertainties except at $z=0$. The local Universe, $z=0$, is also the only epoch for which a significant investigation of this kind has been performed observationally. This is demonstrated by the large differences between the various $z=0$ observational points (black symbols), and in particular in the difference between the \citet{BernardiM_13a} results (black circles) and the other $z=0$ data points. \citet{BernardiM_13a} have performed a careful analysis in which they improved on SDSS background subtraction for massive galaxies, and also integrated the stellar profiles out to larger radii. This resulted in estimates for the {\it total} stellar mass associated with massive galaxies that revised earlier estimates significantly upwards. The Illustris $z=0$ stellar mass function for the `full' masses results is in better agreement with the \citet{BernardiM_13a} mass function with respect to a comparison between the mass function based on our `fiducial' masses inside $r_{\star}$ and the earlier estimates. This may be expected, as the latter are not necessarily integrated out to the same radii between Illustris and observations.

\begin{figure}
\centering
\includegraphics[width=0.475\textwidth]{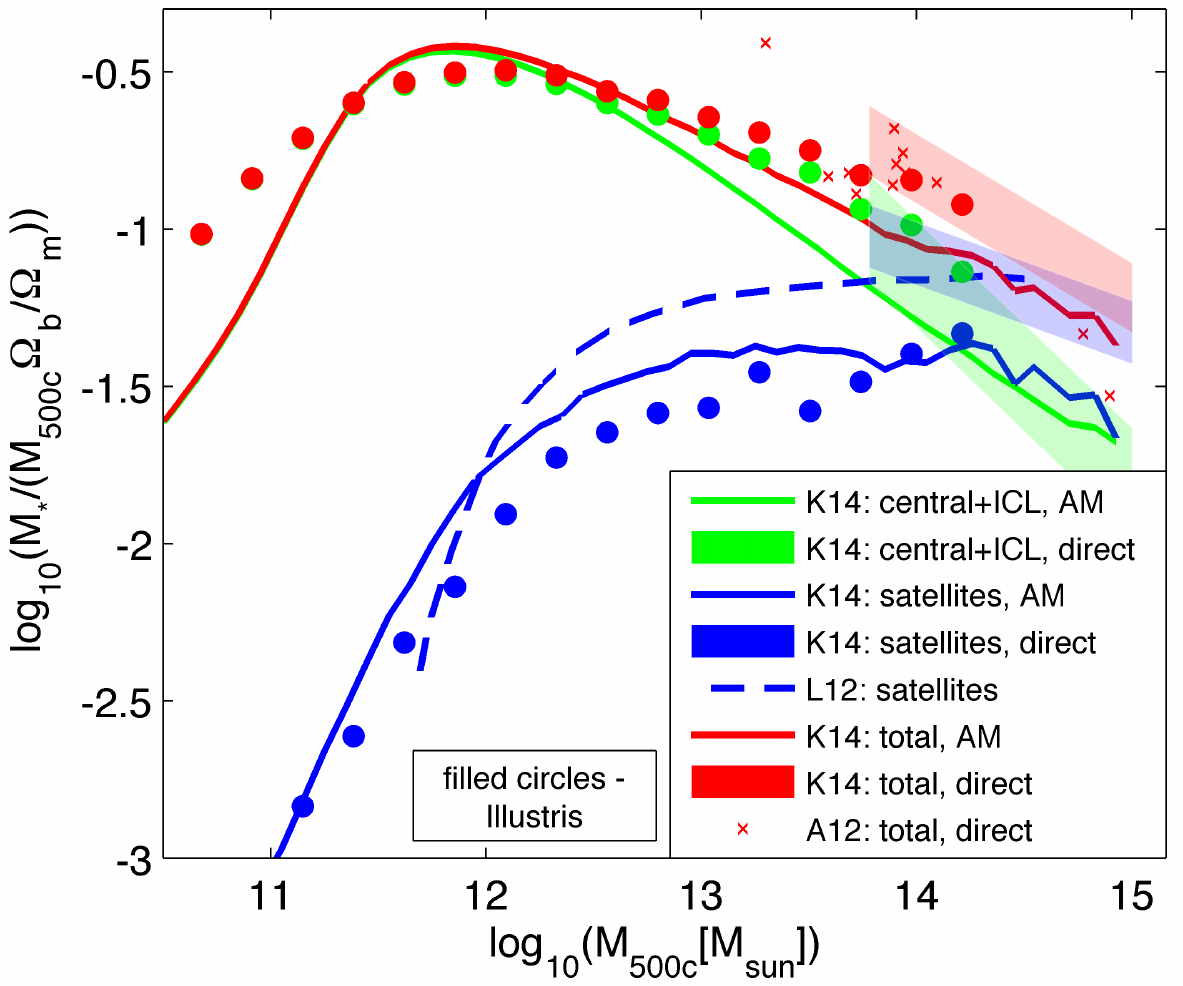}
\caption{Baryon conversion efficiency as a function of halo mass, for two components of the stellar mass -- the central galaxy and its associated mass out to the virial radius $R_{500c}$ (green), and mass contained in satellites that lie within $R_{500c}$ (blue) -- and their sum, which is the total stellar mass within $R_{500c}$ (red). We present results from Illustris (filled circles) alongside relations derived using halo occupation modeling from \citet{KravtsovA_14a} (K14; solid) and \citet{LeauthaudA_12b} (L12; dashed), as well as results based on direct observations of individual halos (shaded regions from K14, and crosses from \citealp{AndreonS_12a}; A12). Focusing on the massive end, we find indication that the slope of the decreasing baryon conversion efficiency toward higher masses is possibly too shallow in Illustris, resulting in central galaxies that are probably somewhat too massive. However, for the more robust measure of total mass, the comparison to direct measurements shows very good agreement with the data.}
\vspace{0.3cm}
\label{f:MsMhz0R500}
\end{figure}

In \Fig{MsMhzallR200Illustris} we show the relation between stellar mass and halo mass for both mass definitions as in \Fig{SMF}. However, a direct comparison to observations that do not consider the full stellar mass will require a significantly more detailed and careful approach in the analysis of simulations to match the procedures used by observers, given the systematics involved with different definitions for the `edge' of galaxies. To circumvent such complications, we show in \Fig{MsMhz0R500} (green) the baryon conversion efficiency at $z=0$ using the `full' stellar masses of the central component (i.e.~excluding satellites) as a function of halo mass (here using $M_{500c}$), and compare to the only available observational results of this kind \citep{KravtsovA_14a}. It is apparent that the baryon conversion efficiency in Illustris (circles) does not decrease with increasing halo mass as strongly as suggested by the abundance matching results (solid), and hence for the most massive halos, the stellar mass is overestimated. However, the \citet{KravtsovA_14a} direct measurements of $21$ clusters (shaded region), and similar data from \citet{AndreonS_12a} (crosses), are in tension with the abundance matching results of the former. The origin of this tension is currently unclear (A.~Kravtsov, private communication), so it is possible that the direct measurements are in fact more faithful to reality than the abundance matching results, in which case the tension with Illustris is virtually gone\footnote{The single data point from \citet{AndreonS_12a} at $M_{500c}\approx10^{13.2}\Msun$ that may appear to be in tension with all the other data, observed and simulated, is in fact not a concern: it is a direct measurement of only a single halo, and the observational uncertainties (not shown here) place it within $1\sigma$ of the Illustris results.}.

Satellite galaxies in groups and clusters do not dominate the galaxy mass function except for at the smallest masses. However, in the most massive halos ($M_{500c}\gtrsim10^{14}\Msun$) they do dominate the total stellar mass over the central component. Also, their properties, such as masses, colours, and morphologies, can be strongly affected by the extreme environments in which they reside. Therefore, they can serve as important probes of galaxy assembly processes. In \Fig{MsMhz0R500} we show the satellite stellar mass as a function of host mass (blue), and compare Illustris to observational inferences by \citet{KravtsovA_14a} and \citet{LeauthaudA_12b}. The `satellite baryon conversion efficiency' is a monotonic function of halo mass, as opposed to the central (green), or total (red), baryon conversion efficiencies. The trend with halo mass is reproduced very well by Illustris. We find that the normalization of the satellite stellar mass is however low in Illustris, by $\approx0.1\dex$ compared to \citet{KravtsovA_14a}, and $\approx0.25\dex$ compared to \citet{LeauthaudA_12b}. Nevertheless, we note that the distinction between the diffuse stellar halo and the satellites is performed in very different ways between the observations (for example, \citet{KravtsovA_14a} integrate a double-S{\'e}rsic profile fit out to infinity to obtain the total luminosity of each satellite galaxy) and the simulation (using saddle points in the density field and a gravitationally-boundedness criterion implemented by \SUBFIND), such that systematic biases in normalization are expected. For the satellites too there exists a tension between the \citet{KravtsovA_14a} abundance matching results and their direct measurements (but note that the halo occupation modeling of \citet{LeauthaudA_12b} agrees well with the direct measurements), so larger samples and a more detailed comparison are required to make definite conclusions regarding agreement between simulations and observations in this respect.

\begin{figure}
\centering
\includegraphics[width=0.475\textwidth]{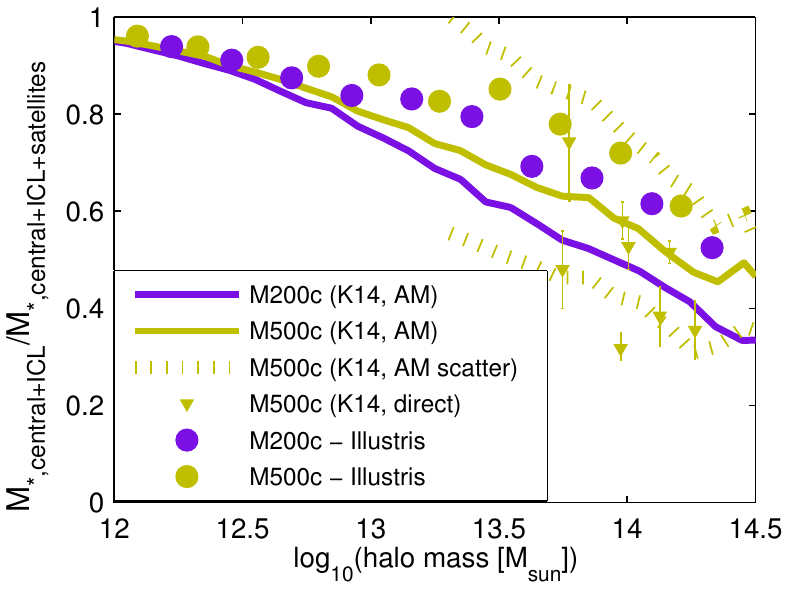}
\caption{The stellar mass fraction associated with the central component of the halo, i.e.~including the central galaxy and diffuse halo stars, with respect to the total stellar mass inside the virial radius, which includes in addition also satellite galaxies. Results are shown for two radii, $R_{200c}$ (purple) and $R_{500c}$ (beige), and based on Illustris (circles), and on \citet{KravtsovA_14a} (K14; curves for abundance matching, and triangles for direct measurements).}
\vspace{0.3cm}
\label{f:BCGfracz0}
\end{figure}

In \Fig{BCGfracz0} we present similar data to that in \Fig{MsMhz0R500} but here the central component is shown as the fraction of the total stellar mass it constitutes in the halo. This is shown for masses inside $R_{500c}$, as in \Fig{MsMhz0R500} (beige), as well as for $R_{200c}$ (magenta). As expected, the central galaxy dominates the more central regions with respect to satellites, such that the fraction associated with it is larger inside $R_{500c}$ than inside $R_{200c}$. The central component in Illustris (filled circles) tends to comprise a larger fraction of the total mass with respect to the abundance matching results (solid curves). However, while it may be subject to biases described above, it is nevertheless within the scatter (dashed curves). In addition, we show estimates from observations of individual clusters (triangles; \citealp{KravtsovA_14a}), which may indicate a further smaller fractional contribution of the central component relative to the abundance matching results.

\subsection{Number density profiles of satellite galaxies}
\label{s:satellites}
In the preceding discussion of satellite galaxies, we have only examined their contribution to the mass budget. However, their spatial arrangement inside galaxy groups and clusters is sensitive to several physical processes that become increasingly significant in these extreme environments, and as such the study of this aspect can strongly constrain galaxy formation models. Gravitationally-induced processes such as tidal stripping, dynamical friction, and merging, as well as hydrodynamical processes such as ram pressure stripping and strangulation that indirectly affect the stellar component by halting further star-formation, all play a role in shaping the spatial distribution of galaxy satellites of different masses inside and in the vicinity of galaxy groups and clusters. The evolution of this spatial distribution is governed by a complicated interplay between processes that occur externally to the groups and clusters, most importantly cosmic structure formation that leads to continuous accretion of new galaxies into these massive objects, and the processes mentioned above that occur mostly inside their deep potential wells.

\begin{figure*}
\centering
\subfigure[]{
          \label{f:satellite_profile_Budzynski}
          \includegraphics[width=0.32\textwidth]{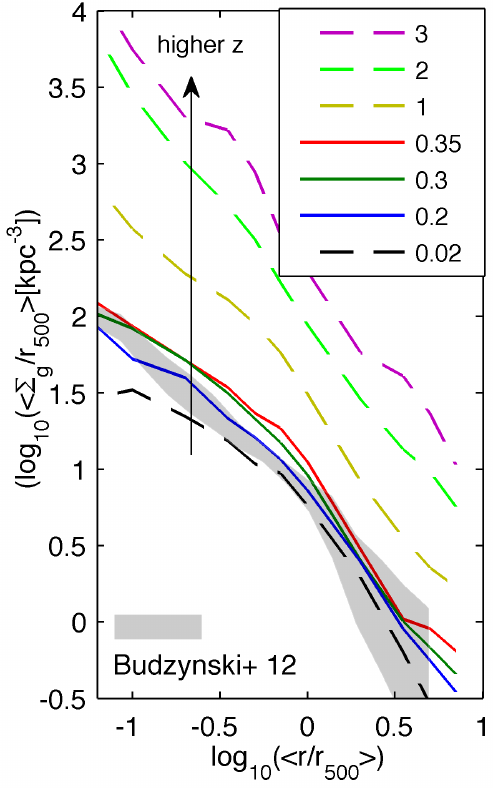}}
\subfigure[]{
          \label{f:satellite_profile_Budzynski_TalUnits}
          \includegraphics[width=0.32\textwidth]{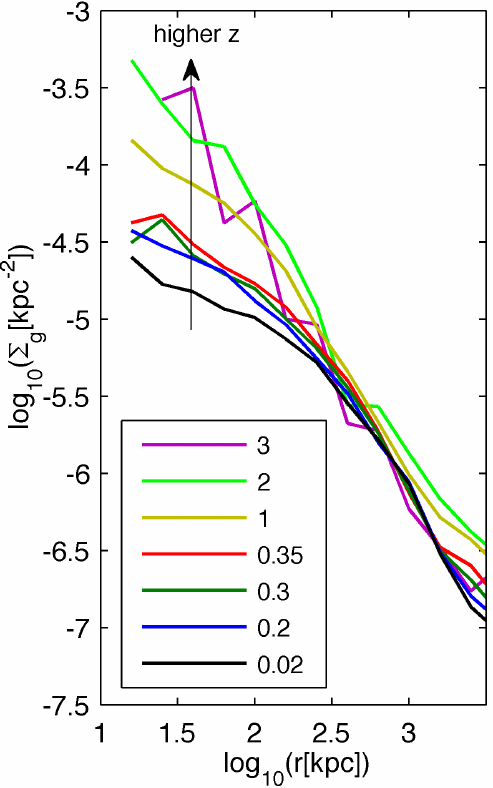}}
\subfigure[]{
          \label{f:satellite_profile_Tal}
          \includegraphics[width=0.32\textwidth]{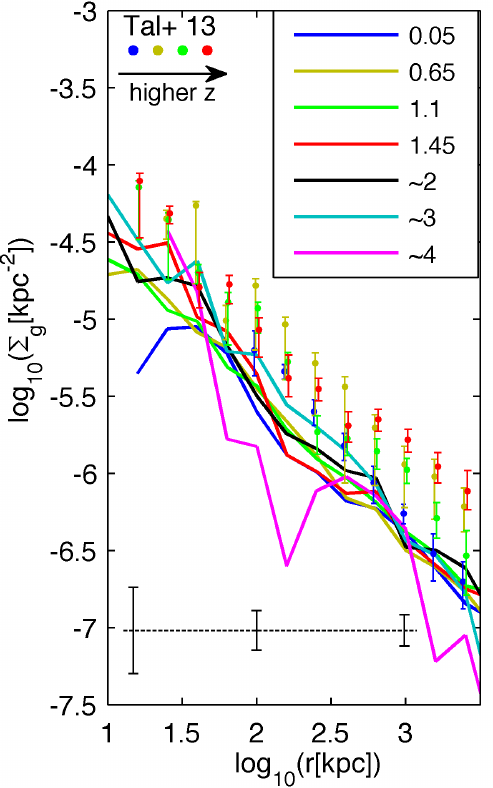}}          
\caption{Radial galaxy number density profiles around massive galaxies and their evolution. {\it Left panel:} The selection criteria for central as well as neighbouring galaxies follow those used observationally by \citet{BudzynskiJ_12a}. The curves extracted from Illustris that correspond to the observed redshift range at $0.15<z<0.4$ (solid curves) show a weak redshift evolution and an excellent agreement with the observations (grey shaded region; compare to Figure 9 in \citealp{BudzynskiJ_12a}). However, Illustris predicts a strong redshift evolution for a larger redshift range (dashed curves). {\it Middle panel:} The same data as in the left panel, when plotted in physical units rather than in units normalised to $R_{500c}$, shows a much reduced redshift evolution, and is in particular consistent with no evolution at $r\gtrsim300\kpc$. {\it Right panel:} The selection criteria follow \citet{TalT_13a}, with the neighbouring galaxies relating to the central ones by a stellar mass ratio of $1:10$. The $z=0.05, 0.65, 1.1, 1.45$ Illustris curves match to the observations of \citet{TalT_13a}, which are shown by the symbols with error bars. With this selection, and presented in physical units, observations show no significant redshift evolution for $0\lesssim z\lesssim1.5$. Illustris reproduces this non-evolution, and predicts that it continues up to $z\approx4$. The three error bars at the bottom of the panel demonstrate the typical magnitude of the Poisson errors of the simulation results at three representative radii, averaged over the various redshift bins.}
\vspace{0.3cm}
\label{f:satellite_profiles}
\end{figure*}

In \Fig{satellite_profiles} we compare results from Illustris to two independent measurements of galaxy number density radial profiles around massive galaxies at various redshifts. We begin with a comparison shown in \Fig{satellite_profile_Budzynski} to the observations made by \citet{BudzynskiJ_12a} (grey shaded region). These authors used SDSS to select central galaxies of halos at $0.15<z<0.4$ with $10^{13.7}<M_{500c}[\Msun]<10^{15}$, and measured the radial profiles of galaxies around them that are brighter than $-20.5$ magnitudes in the r-band. When dividing their sample to three independent redshift bins, \citet{BudzynskiJ_12a} found no significant evidence for redshift evolution of the radial profiles. We conducted mock observations of Illustris by measuring the corresponding two-dimensional profiles around the central galaxies of halos of $M_{500c}>10^{13}\Msun$ integrated through the simulation volume, and performing a statistical background subtraction based on random lines of sight through the box, in a way that mimics the \citet{BudzynskiJ_12a} observations (we have, however, neglected dust attenuation). We find that their results are reproduced in Illustris (solid curves) to an excellent level of accuracy (with differences not exceeding $0.1\dex$, or about $1\sigma$; see also \citealp{VogelsbergerM_14a}). It is worth noting that previous theoretical studies, both semi-analytical models and hydrodynamical simulations, have had trouble reproducing similar observations. Semi-analytical models, despite their higher level of freedom in tuning prescriptions for relevant physical processes, such as tidal stripping and dynamical friction, consistently find too many satellite galaxies in the inner parts of clusters \citep{QuilisV_12a,WangW_14a}, while hydrodynamical simulations tend to find radial profiles that are too shallow at small radii \citep{NagaiD_05a,SalesL_07a}. We note that a comparison to Illustris-2 (not shown here) shows that the results in \Fig{satellite_profiles} are extremely well converged with respect to resolution.

Both observations and Illustris see no significant redshift evolution inside this observed redshift range of $0.15<z<0.4$. However, Illustris allows us to extend the estimate of this quantity to both lower and higher redshifts, thereby providing predictions for future observations (dashed curves). Remarkably, we find that the redshift range probed by \citet{BudzynskiJ_12a} is special in showing no significant redshift evolution. The radial galaxy profiles in Illustris show significant evolution from $z\approx0.2$ to $z\approx0$, especially in the inner parts within $R_{500c}$. Moreover, the evolution at $z>0.35$ is rapid, reaching approximately two orders of magnitude higher between $z\approx0.35$ and $z\approx3$, at all radii. We find that outside of $R_{500c}$, the profiles scale as $\Sigma_{\rm gal}\propto r^{\approx-1.5}$. This slope extends well into $R_{500c}$ at $z\gtrsim1$, while at $z\lesssim1$ it flattens inside $R_{500c}$ and becomes $\Sigma_{\rm gal}\propto r^{\approx-1}$. At all redshifts, the profiles at large radii are shallower than the NFW \citep{NavarroJ_97a} profile, while at small radii they are steeper than it. Observing such profiles at redshift ranges that go beyond the \citet{BudzynskiJ_12a} measurement will be a very interesting test of Illustris.

We can gain some insight into the nature of this strong redshift evolution by changing the units of the plot. \citet{BudzynskiJ_12a} normalised both axes by presenting the results in physical units divided by an estimate of $R_{500c}$. In \Fig{satellite_profile_Budzynski_TalUnits} we show the same data as in \Fig{satellite_profile_Budzynski}, but now instead with physical units (${\rm kpc}$) on both axes. The results now appear strikingly different. The redshift dependence is almost entirely gone at large radii, with all profiles at $r\gtrsim300\kpc$ being statistically indistinguishable. While the halos are selected to have $M_{500c}>10^{13}\Msun$ at each of the redshift bins, and so are not necessarily direct progenitors and descendants of one another, the mass rank-ordering does not change dramatically for these high masses and this redshift range \citep{LejaJ_13a}. Therefore, we can interpret this finding as a result of a roughly constant inwards flow of bright galaxies inside $300\kpc\lesssim r\lesssim3000\kpc$, such that the number density remains constant with time. However, there remains a significant downwards evolution of the profiles at $r\lesssim300\kpc$, meaning that bright galaxies at these small radii are either stripped or merged at a higher rate than they are provided from larger radii by cosmological accretion.

Next, we perform a detailed comparison to the observations made by \citet{TalT_13a}, which span a much larger redshift range. \citet{TalT_13a} selected massive galaxies with a fixed number density of $4\times10^{-5}\pMpccb$ in an attempt to identify populations of progenitors and descendants, in a statistical sense. Also the selection of neighbouring galaxies differs somewhat from what was done in \citet{BudzynskiJ_12a}; here, the neighbouring galaxies are chosen with a fixed stellar mass ratio ($1:10$) with respect to the primaries, rather than a fixed luminosity. \citet{TalT_13a} quantified number density profiles from $z\approx0$ up to $z\approx1.6$, and found no statistically significant redshift evolution. We repeat their procedure with mock observations of Illustris, and in agreement with their results, we find no redshift evolution, which extends, with this selection criteria, down to the smallest radii\footnote{The apparent drop of the $z=0.05$ curve at $r\lesssim30\kpc$ is probably driven by Poisson noise of the very small number of satellites in these inner bins ($\approx2$). Illustris-2 shows no such inner drop.}, as opposed to the case shown in \Fig{satellite_profile_Budzynski_TalUnits}. We also extend the measurement to much higher redshifts, up to $z=4$, and find that the constancy of the profiles with redshift continues to hold. The normalization we find in Illustris is, however, up to $\approx0.3\dex$ lower than reported by \citet{TalT_13a}. This can be related to the selection being made here based on the mass ratio, rather than a fixed satellite luminosity or mass, since the mass of central galaxies in Illustris is overestimated more than the mass of satellites (see \Figs{MsMhz0R500}{BCGfracz0}), such that their masses are more disparate than in the real Universe. When we repeat the measurement but use a mass ratio of $1:15$ instead of $1:10$, we find a closer agreement with \citet{TalT_13a} (not shown), as expected based on this argument. We interpret this as an indication that the satellite galaxies per se are in very good agreement with the observations, while the central galaxies are somewhat too massive, as we have shown already. We find that the profiles have a constant slope with $\Sigma_{\rm gal}\propto r^{\approx-1}$ at all radii, which is shallower in the outer parts than what is found in \Fig{satellite_profile_Budzynski_TalUnits}. These differences illustrate that satellite profiles can be sensitive to the details of the selection, and that the interpretation of their concentration as compared to the concentration of the underlying DM halo, where different observational works reached opposite conclusions \citep{BudzynskiJ_12a,WatsonD_12a,WojtakR_13a,TalT_13a}, may benefit significantly from comparisons to simulations that go beyond N-body only.

\section{Gas and baryon content of galaxies and halos}
\label{s:gas}
\subsection{Baryon content of halos}
\label{s:gas_halos}
\begin{figure}
\centering
\includegraphics[width=0.475\textwidth]{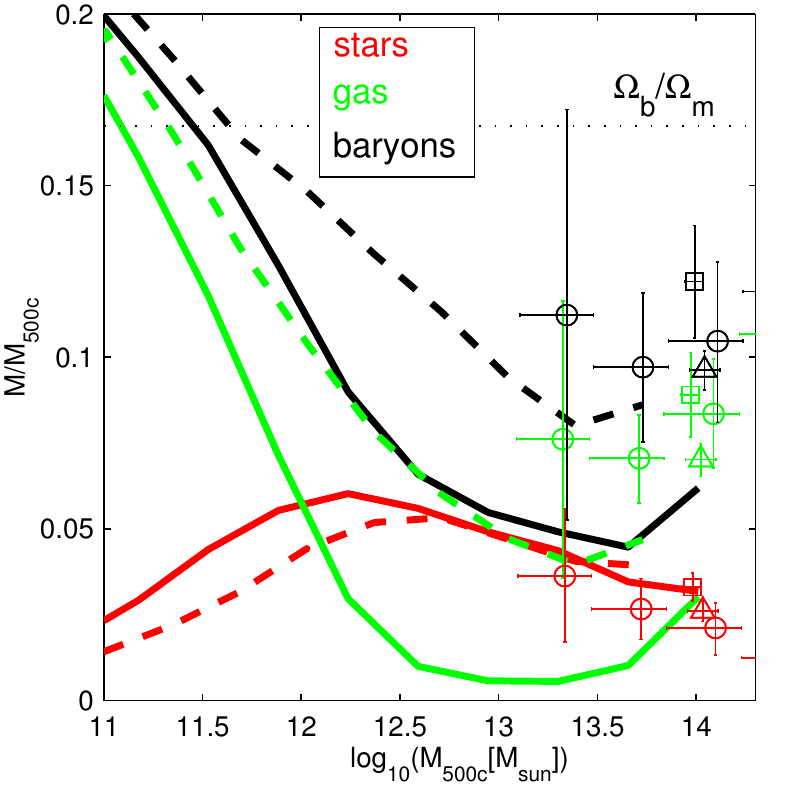}
\caption{Total baryonic mass fractions inside $R_{500c}$ as a function of halo mass (black), separated into contributions from stars (red) and gas (green). Mean relations from Illustris are shown for $z=0$ (solid) as well as for $z=1$ (dashed). For comparison, open symbols show $z\approx0$ observations by \citet{GiodiniS_09a} (circles), \citet{GonzalezA_13a} (squares), and \citet{SandersonA_13a} (triangles). The stellar masses match observations rather well, as discussed in Section \ref{s:z0MassiveEnd}, however gas masses for massive halos at $z=0$ are severely underestimated by the simulation.}
\vspace{0.3cm}
\label{f:MbMhz01R500}
\end{figure}

Galaxies, and even more so their host halos, can contain a substantial fraction of their mass as diffuse gas. The gas can be multi-phase, with hot and ionized gas filling most of the volume of DM halos, and dense, cold star-forming gas, whether neutral or molecular, building galactic gas disks. The different phases interact and exchange mass, and are also intimately connected to black holes and stars via accretion, star-formation, and feedback processes. Therefore, studying the gas properties in cosmological simulations can both constrain theoretical and numerical models, and serve as a useful tool for understanding the cosmic baryon cycle. In \Fig{MbMhz01R500} we show the mean fraction of the mass in the form of stars (red), gas (green), and all baryons (black), as a function of halo mass, for $z=0$ (solid) and $z=1$ (dashed), as derived from Illustris, and compare to $z=0$ observations (symbols with error bars). As discussed in Section \ref{s:z0MassiveEnd}, the total stellar mass fractions of massive halos in Illustris at $z=0$ are quite close to those observed, and are a decreasing function of halo mass. As for gas mass measurements, these masses of $M_{500c}\approx10^{13}-10^{14}\Msun$ are just high enough for observational constraints from hot atmosphere modeling to become available, but lower than the most massive clusters (which we lack due to a limited box size) where the observations indicate an increasing baryon fraction with increasing halo mass, up to the cosmic baryon fraction. The observed relation for this mass range is flat, with a slight indication for an increase above $M_{500c}\approx10^{13.8}\Msun$. In Illustris, the gas (and total baryon) fraction is indeed flat in this mass range (and begins to rise above $M_{500c}\approx10^{13.5}\Msun$), since this is the regime in which the radio-mode feedback is most effective: towards lower masses, it becomes less effective as the black holes are less massive (Sijacki et al.~in prep.), and towards higher halo masses, it becomes less effective as the potential wells become deeper. However, the total baryonic mass fractions, as well as the gas mass fractions that dominate the total baryon content, are underestimated in Illustris with respect to observations. The gas masses at $M_{500c}\approx10^{13}\Msun$ are underestimated by a factor of $\approx10$, and at $M_{500c}\approx10^{14}\Msun$ by a factor of $\approx3$. This likely indicates that the radio-mode feedback, whose strength was tuned to suppress star-formation and stellar mass buildup in these massive halos, operates in a way that is too violent for the gas component, ejecting large gas masses out of the halos. We note that \Fig{MbMhz01R500} also shows that the simulated gas, and baryonic, fractions at $z=1$ are significantly higher than at $z=0$. This is a result of the late-time onset of the efficient radio-mode feedback, whose role in suppressing star-formation within our model becomes increasingly important at lower redshifts.

It is interesting to note that as the radio-mode AGN feedback becomes less effective towards low masses below $M_{500c}\approx10^{12.5}\Msun$, we find rapidly increasing halo gas fractions that approach the cosmic baryon fraction. This is understandable as the velocity that is assigned to the galactic winds in our model scales with, but is somewhat lower in magnitude than, the escape velocity from the halo to infinity. As a result, the galactic winds generally do not eject gas out of halos, allowing them to retain their cosmic baryon fractions (Suresh et al.~in prep.). Such large gas reservoirs have not yet been detected around $L^*$ or fainter galaxies. However, this may be due to observational uncertainties and limitations \citep{AndersonM_14a}, which are currently only sensitive to the inner parts of the halos \citep{BogdanA_12a,BogdanA_13a,AndersonM_13a}. The observed limits are quite sensitive to model assumptions \citep{AndersonM_10a}, such that a far more detailed modeling of the exact conditions of this predicted gas is needed in order to give faithful predictions for its observability. Halos with $M_{500c}\approx10^{11}\Msun$ in Illustris are found to have a baryon fraction even somewhat above the cosmic baryon fraction, which is related to the finding that these halos become more massive with respect to their Illustris-Dark counterparts \citep{VogelsbergerM_14b}. The baryon fraction, however, starts falling towards lower masses below the peak at $M_{500c}\approx10^{10.7}\Msun$ (not shown), under the influence of the cosmic ionizing background radiation \citep{OkamotoT_08a}.

\begin{figure*}
\centering
\includegraphics[width=1.0\textwidth]{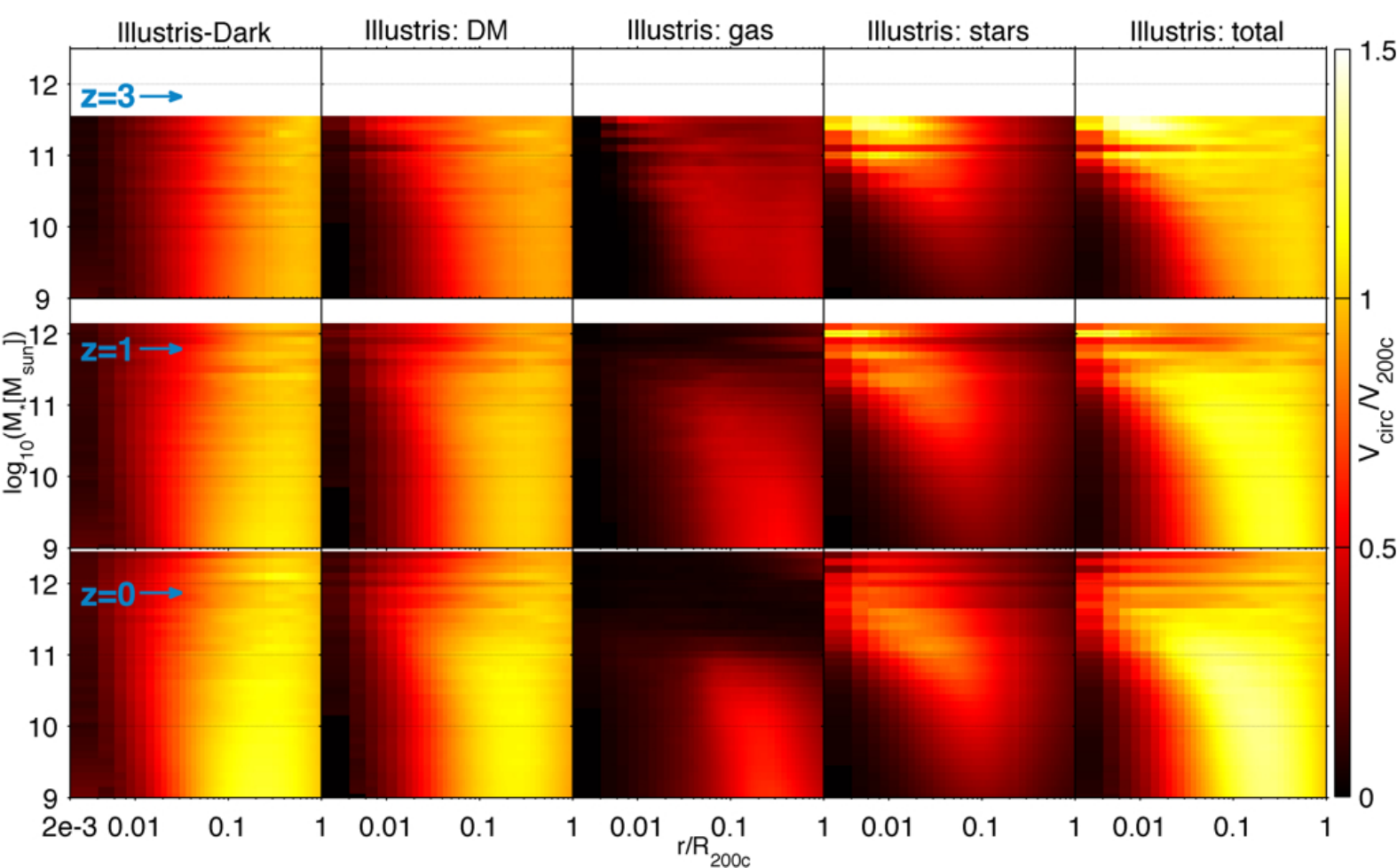}
\caption{Circular velocity profiles for central galaxies as a function of radius, stacked by stellar mass. The colour indicates the mean circular velocity, normalised by the mean $V_{200c}$ of the corresponding host halos. The different rows are for different redshifts. The left-most column shows the results from Illustris-Dark, the DM-only version of the Illustris volume. The other columns show the contribution of different mass components $\sqrt{GM_i(<r)/r}$ to the total circular velocity in the full-physics Illustris Simulation, where $i$ stands for: DM, gas, or stars. The right-most column shows the total circular velocity in Illustris, $\sqrt{GM(<r)/r}$. The white `gaps' at the highest masses in the $z>0$ panels are a result of the lack of galaxies with such high stellar masses at those redshifts.}
\vspace{0.3cm}
\label{f:circular_velocity_components}
\end{figure*}

In \Fig{circular_velocity_components} we go beyond quantifying the total baryonic content of halos, and look into the radial distribution of various mass components. For three different redshifts, $z=3$, $z=1$, and $z=0$, we show the total circular velocity profile $\sqrt{GM(<r)/r}$ for central galaxies (right column), normalised by $V_{200c}$, as a function of radius and stacked by stellar mass. The three middle columns break the total circular velocity profile into its contributions coming from DM, gas, and stars, and the left panel shows the circular velocity profile for the matched halos from Illustris-Dark. 

We begin the discussion with the $z=0$ results (bottom row). A comparison between the two panels on the extreme sides reveals the effects of baryonic physics, most of all baryon condensation, on the total mass distribution inside halos. \citet{VelliscigM_14a} have recently shown that the total enclosed mass profiles of halos are sensitive to the implementation of baryonic physics and feedback. Indeed, we find that halos hosting galaxies with stellar mass $M_*\lesssim10^{11}\Msun$ have $V_{\rm max}/V_{200c}\approx1.1$ in Illustris-Dark, however in Illustris, where baryonic physics is included, the circular velocity curve is more peaked, with $V_{\rm max}/V_{200c}\approx1.3$. This is an important probe of galaxy formation processes, which is relevant for obtaining a simultaneous match to the stellar mass function and the Tully-Fisher relation \citep{VogelsbergerM_14b}, and for which we find good agreement with observations \citep{ReyesR_12a,CattaneoA_14a}. Moreover, the peak is reached at smaller radii for galaxies with $M_*\lesssim10^{11.5}\Msun$, at $r\approx(0.05-0.2)R_{200c}$, compared with its occurrence at $r\approx0.3R_{200c}$ in Illustris-Dark. Importantly, however, the peak tends to occur at radii larger than a few percent of the virial radius, and at the higher masses where it is closer to the center, it is also very broad. This means that the rotation curves of galaxies in Illustris are rising and are thereafter flat, and only start decreasing at a large fraction of the virial radius, in good agreement with observations.

An examination of the third and fourth panels from the left reveals that for galaxies with stellar mass $M_*\lesssim10^{10}\Msun$, the contracted profiles with respect to Illustris-Dark are dominated by a significant gas contribution at $r\approx0.2R_{200c}$, while for galaxies with stellar mass $10^{10.5}\Msun\lesssim M_*\lesssim10^{11.5}\Msun$, the baryon concentration is mostly in the form of stars at $r\approx(0.01-0.1)R_{200c}$. A comparison between the first two panels from the left shows that the baryons affect the distribution of DM itself in galaxies with $M_*\gtrsim10^{10.5}\Msun$ (see also \citealp{DiCintioA_14a}). In particular, in the mass range $10^{10.5}\Msun\lesssim M_*\lesssim10^{11.5}\Msun$ there is clear evidence for contraction of the DM halo in response to the central stellar concentration at $r\approx(0.01-0.1)R_{200c}$. In the most massive galaxies, with $M_*\gtrsim10^{11.5}\Msun$, a third regime is found, where the contribution of the stars to $V_c/V_{200c}$ drops significantly below unity (as the stellar component becomes significantly more extended), the contribution of gas becomes negligible (as a result of radio-mode AGN feedback), the DM is not contracted any more in response to the baryons (see also \citealp{MeadJ_10a}), and so the total circular velocity profiles become only very slightly modified with respect to the DM-only simulation Illustris-Dark. Detailed studies of the profiles and shapes of DM halos as well as stellar halos in Illustris are presented elsewhere (\citealp{PillepichA_14a}, Chua et al.~in prep.).

At $z>0$, this general picture still holds, however important differences do exist. Most notably, massive galaxies with $M_*\gtrsim10^{11}\Msun$ have stellar concentrations that are much stronger towards higher redshifts, dominating maximum circular velocities at $z\sim3$ that are in excess of $1.5V_{200c}$ at radii of $\approx0.01R_{200c}$. Moreover, the gas component, which becomes negligible at $z=0$ for $M_*\gtrsim10^{11}\Msun$, plays a more important role at higher redshifts, becoming more concentrated at higher masses in a way that approximately follows the stellar component. As a result, the DM is contracted with respect to Illustris-Dark up to the most massive galaxies in the box. For example, at a scale of $M_*\approx10^{11.5}\Msun$, where the effect at $z=0$ has started to reverse already from maximum contraction to expansion, due to the operation of the radio-mode AGN feedback, galaxies at $z=3$ show significant contraction. For lower mass galaxies, it is not the stars but the DM that sets the peak of the circular velocity profile. The peak for those galaxies has a lower amplitude than at low redshift, since higher redshift \LCDM{ }halos are less concentrated \citep{MunozCuartasJ_11a}, as seen in the left-most panel. The combination of these two trends means that the $V_{\rm max}$-$M_*$ relation becomes steeper towards higher redshifts, which may reveal itself observationally as a steepening of the Tully-Fisher relation (depending on the relation between $V_{\rm max}$ and the rotation velocity, or other probes used for this purpose, and the radius at which they are measured).

\subsection{Gas content of galaxies}
\label{s:gas_galaxies}
We now turn to study gas masses on the galactic scale. In \citet{VogelsbergerM_14a} we demonstrate good agreement with observations for the HI richness of $z\sim0$ galaxies. To complement that comparison, and since there are no available measurements for neutral gas at higher redshifts, we focus here on the molecular gas phase. Observationally, the molecular gas fraction is found to be larger for less massive galaxies, and to increase towards higher redshifts, at least up to $z\sim2$.

Before making a comparison to the simulation results, it should be noted that the data on molecular gas is subject to systematic uncertainties both on the observational and on the theoretical sides. For example, the observations we compare to here are based on CO measurements, which must rely on assumptions for the CO-to-H2 conversion factor to derive total molecular masses \citep{GenzelR_12a}. Also, data at $z>0$ are still scarce due to the high sensitivity required to make such measurements, and suffer from selection effects.

On the simulation side, where we do not explicitly follow the various cold gas phases of the interstellar medium, we need to make assumptions as for what constitutes the molecular phase. While several empirical and theoretical prescriptions (\citealp{BlitzL_06a,KrumholzM_09b}; see also \citealp{SternbergA_14a}) have been adopted in various semi-analytical or hydrodynamical simulations for partitioning the cold gas into the neutral and molecular phases \citep{FuJ_10a,LagosC_11a,AltayG_11a,CenR_12a}, these prescriptions are not well-calibrated against data at high redshift, and do not agree among themselves in various regimes. Therefore, we choose here to take a simple approach. Results from the simulation are based on the star-forming gas inside the fiducial galactic radius $r_{\star}$. For high redshift, the underlying assumption that most of the star-forming gas is in the molecular phase is based on the large column densities typical for those redshifts \citep{ObreschkowD_09a}. For $z\sim0$, a substantial fraction of the star-forming, cold gas is in the neutral phase, hence the full star-forming gas mass inside the galactic radius is expected to represent an over-estimate of the molecular gas mass. To account for that, we use the empirical finding that the molecular-to-neutral mass ratio at $z\sim0$ is approximately $1/3$ \citep{SaintongeA_11a} to derive a lower limit for the simulated molecular mass by dividing the star-forming gas mass by a factor of $4$. This is expected to be a lower limit, since the observed molecular-to-neutral mass ratio is based on the total neutral gas, including gas at large radii, such that the molecular-to-neutral mass ratio inside the radius we consider should be larger than $1/3$.

\begin{figure}
\centering
\includegraphics[width=0.475\textwidth]{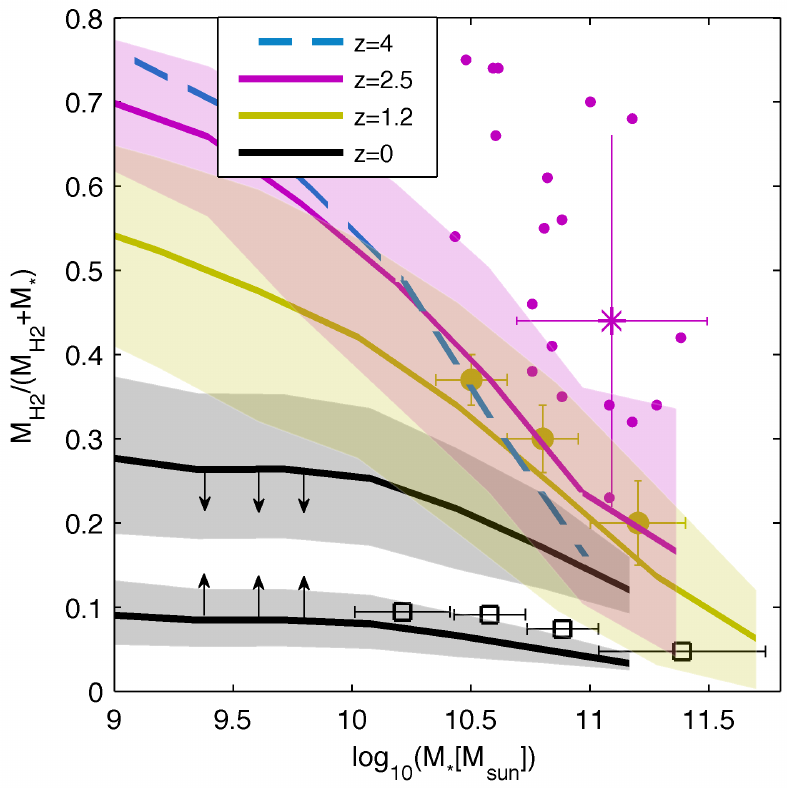}
\caption{Galactic molecular gas fractions as a function of stellar mass, where results from Illustris (curves with $1\sigma$ scatter represented by shaded regions) are compared to observations. Observational data points at $z\sim0$ (squares) are adopted from \citet{TacconiL_13a}, based on the COLD GASS survey \citep{SaintongeA_11a}, and only include galaxies within $0.5\dex$ of the star-formation main sequence. Observational data at $z>0$ were collected as part of the PHIBSS survey, and are adopted from \citet{TacconiL_10a} (asterisk) and \citet{TacconiL_13a} (circles). The symbols without error bars represent individual objects, and the symbols with error bars represent galaxy samples. The error bars in the horizontal direction denote the bin size, and the ones in the vertical direction give the uncertainty in the mean. For the simulation results at $z=0$, we provide two estimates that are expected to correspond to upper and lower limits (see text for details), and also match the observational selection of galaxies around the mean sequence (hence the black curves do not extend as far in stellar mass as the higher-redshift curves).}
\vspace{0.3cm}
\label{f:gas_fraction}
\end{figure}

With the above assumptions regarding what constitutes the molecular gas mass of galaxies $M_{\rm H2}$, we show in \Fig{gas_fraction} the mean molecular gas fraction ($M_{\rm H2}/(M_{\rm H2}+M_*)$) as a function of stellar mass, for several redshifts. We identify several trends in Illustris (curves with $1\sigma$ scatter represented by shaded regions), and compare them to observations. First, we find that lower-mass galaxies have larger molecular gas fractions. The slope of the relation with stellar mass matches the observations very well at all redshifts. We also find that the relation becomes steeper at higher redshifts, which for $z\gtrsim2$ stands in sharp contrast to the prediction of the semi-analytical model by \citet{PoppingG_14a}, where the dependence on mass becomes weaker at higher redshifts. Second, we find that the relation evolves towards higher gas fractions at higher redshifts, up to $z\approx2$. At $z=0$, our two estimates for the normalization of the relation, which represent upper and lower limits from the simulation, bracket the data points. At $z\approx1.2$, the simulation results are approximately $1\sigma$ lower than the observations, a difference that is smaller than the scatter and the systematic uncertainties. Therefore, we conclude that we obtain a good match at this redshift. At $z\approx2.5$, the mean gas fraction in the simulation is a factor of $\approx2$ lower than inferred from observations, while the trend with mass and the scatter match well. At $z\gtrsim2$, the gas fractions generally stop increasing, and remain approximately constant at least out to $z\approx4$ (albeit with some mass-dependent changes, as a result of the aforementioned steepening). Such a qualitative behaviour has been suggested in recent $z\sim3$ data \citep{MagdisG_12a,SaintongeA_13a}, and similar limits have been set even at $z\sim5$ \citep{DaviesL_10a,LivermoreR_12a}, although large uncertainties still plague the measurements at such high redshifts. It is possible that the relation stops evolving at too low a redshift in Illustris with respect to observations, making the $z\approx2.5$ gas fractions too low. This weak redshift evolution at $z\gtrsim2$ is at odds with the prediction of the semi-analytical model by \citet{PoppingG_14a}, where the clear redshift evolution continues well above $z\sim2$. New observational data is expected to be able to reliably distinguish between these different predictions for $z\gtrsim3$ in the next few years \citep{ObreschkowD_11a}, both for the global normalization evolution and for the evolution of the slope as a function of stellar mass.

\section{Star-formation rates}
\label{s:sSFR}
Closely related to the gas fraction of galaxies is their specific star-formation rate (sSFR). The relation is however physically complicated, and often parameterised using the notion of the star-formation efficiency or timescale. The local star-formation timescale in our model \citep{SpringelV_03a} is a function of gas density, hence the overall star-formation timescale of a galaxy depends on the density distribution of its star-forming gas, which in turn is determined by dynamical, as well as feedback, processes. Recent theoretical models suggest that it may be useful to think of the star-formation rate as a more fundamental quantity than the gas fraction \citep{BoucheN_10a,DaveR_12a,DekelA_14a}. In these models, the star-formation rate is set by the existence of a mass-conserving baryon cycle of cosmological accretion and gas cooling on one hand, and ejection by feedback on the other hand. The specific accretion rate of DM halos (sDMAR) can be derived from N-body simulations \citep{GenelS_08a} to follow simple power-law relations with mass and redshift, ${\rm DMAR}\propto M^{\approx0.1}(1+z)^{\approx2.2}$, to a good level of approximation (in reasonable agreement with analytical derivations, \citealp{BirnboimY_07a}). Indeed, the gas supply rate onto a galaxy is governed by physics that is distinct from the gravitationally-induced growth of its surrounding halo \citep{NelsonD_13a}. Nevertheless, observations indicate similar sSFR dependencies on mass and redshift to the sDMAR. 

The challenge to galaxy formation theories lies in explaining the few notable differences between the two that do exist. These differences are the following: {\bf (a)} a normalization offset, such that the sSFRs are higher than the sDMARs by factors of $\approx3-10$, {\bf (b)} the mass dependence is weak in both cases, but inverted, such that high-mass halos have higher sDMARs than their less massive counterparts, but more massive galaxies have lower sSFRs than their low-mass counterparts, and {\bf (c)} the redshift dependence of the sSFR evolves faster at $z\lesssim2$, and then possibly slower at $z\gtrsim2$ with respect to the sDMARs.

\begin{figure}
\centering
\includegraphics[width=0.475\textwidth]{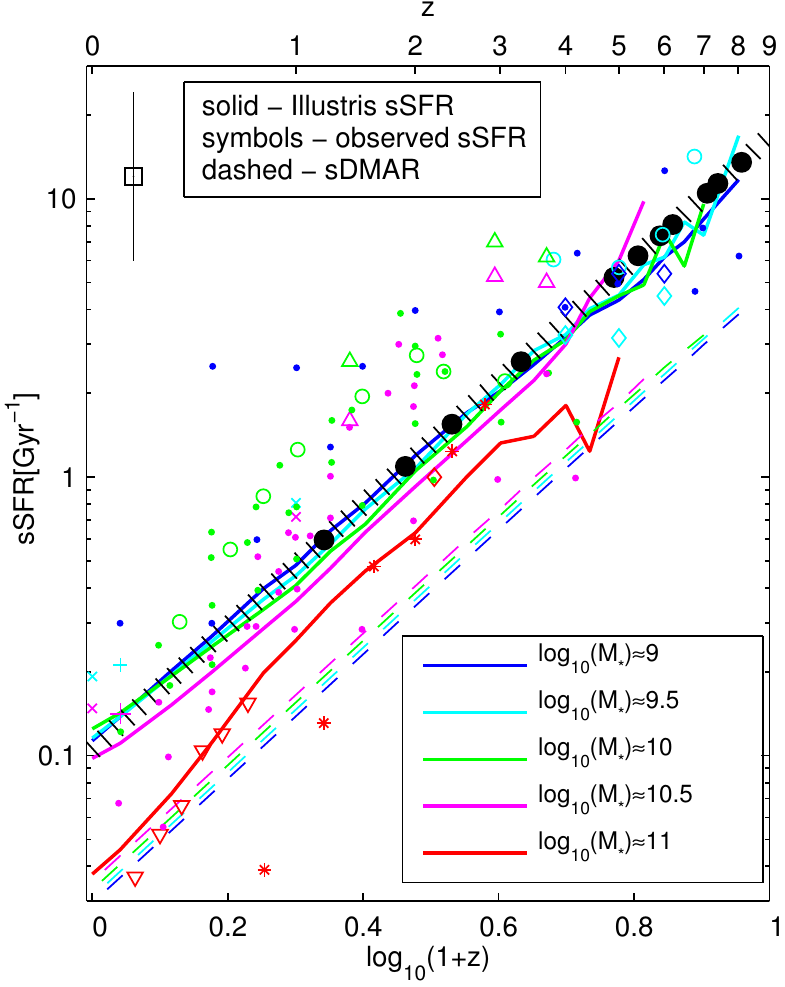}
\caption{Specific star-formation rate ($\SFR/M_*$) in Illustris as a function of redshift for five $0.5\dex$-wide mass bins centered around $M_*=10^{9},10^{9.5},10^{10},10^{10.5},10^{11}\Msun$ (solid curves). The specific mass accretion rate of the DM halos hosting the galaxies in each bin is shown by a corresponding dashed curve, based on the fitting formula provided by \citet{GenelS_08a}. The dashed black curve shows the specific mass accretion rate of DM halos of $M_h=10^{12}\Msun$, multiplied by a constant factor of $3$. For comparison, observational data points are adopted from the compilation by \citet{BehrooziP_13b} (filled circles), and augmented with the observations of \citet{NoeskeK_07a} (green open circles), \citet{ElbazD_07a} (crosses), \citet{SalimS_07a} (pluses), \citet{DamenM_09a} (red asterisks), \citet{OliverS_10} (red triangles), \citet{StarkD_13a} (cyan open circles), \citet{GonzalezV_14a} (diamonds), and \citet{Heinis_14a} (upward triangles). The observational error bars have been omitted for visual clarity, and instead the typical uncertainty of a factor of $\approx2$ is represented by the black error bar in the top-left.}
\vspace{0.3cm}
\label{f:sSFR_vs_z}
\end{figure}

In \Fig{sSFR_vs_z} we show the sSFRs of galaxies in Illustris as a function of redshift, for several mass bins (solid curves), and compare to observations (symbols with error bars). We generally find a good agreement, except at $z\sim1-3$, where observations indicate higher values by a factor $\approx2$, and in particular for lower masses, where the disagreement persists also at $z<1$ (see also \citealp{TorreyP_14a}). We also compare to the theoretical sDMARs in which the galaxies reside (dashed curves), which evidently have a similar shape to the sSFRs in the simulation. By multiplying the sDMAR of $10^{12}\Msun$ halos by a factor of $3$ (dashed black curve), we directly show that the two quantities follow each other closely in the simulation, albeit with a normalization offset -- an offset that is required to obtain a close match to the observations. Examining different masses in more detail reveals that Illustris produces sSFRs that are smaller for galaxies with higher stellar masses, in an opposite trend as applies for the sDMARs, which is in qualitative agreement with observations. However, the mass dependence is quantitatively weaker than observed, as for galaxy masses $M_*\lesssim10^{10.5}$, the scaling in Illustris is ${\rm sSFR}\propto M^{\approx-0.08}$, while various observations indicate a large spread of values, ${\rm sSFR}\propto M^{-0.1}-M^{-0.7}$ (that is, however, probably dominated by systematics; \citealp{SpeagleJ_14a}). The most massive galaxies show a significantly stronger mass dependence, as quenching by AGN feedback becomes effective, in possible agreement with some observations (\citealp{OliverS_10}, but see \citealp{DamenM_09a}).

The biggest challenge in this context lies in the redshift dependence, namely in that the $z\sim1-3$ sSFRs in Illustris are lower than observed, while the agreement at both higher and lower redshift is good. This difference in shape can be described in terms of a steep observed rise from $z=0$ to $z\approx2$, that is then followed by a flattening at $z\gtrsim2$, which appears neither in the simulation, nor in the theoretical sDMAR evolution. This qualitative difference is in common to most hydrodynamical simulations, semi-analytical, and semi-numerical models \citep{DamenM_09a,BoucheN_10a,FirmaniC_10a,DaveR_11a,LillyS_13a,DekelA_14a}, with possible solutions being rather radical or contrived \citep{DaveR_08a,KangX_10a,WeinmannS_11a,BirrerS_14a}, leaving the issue as an open challenge for galaxy formation theories.

\section{Galaxy morphologies}
\label{s:morphologies}
\begin{figure*}
\centering
\subfigure[Gas surface density maps]{
          \label{f:gas_images_lowmass}
          \includegraphics[width=1.0\textwidth]{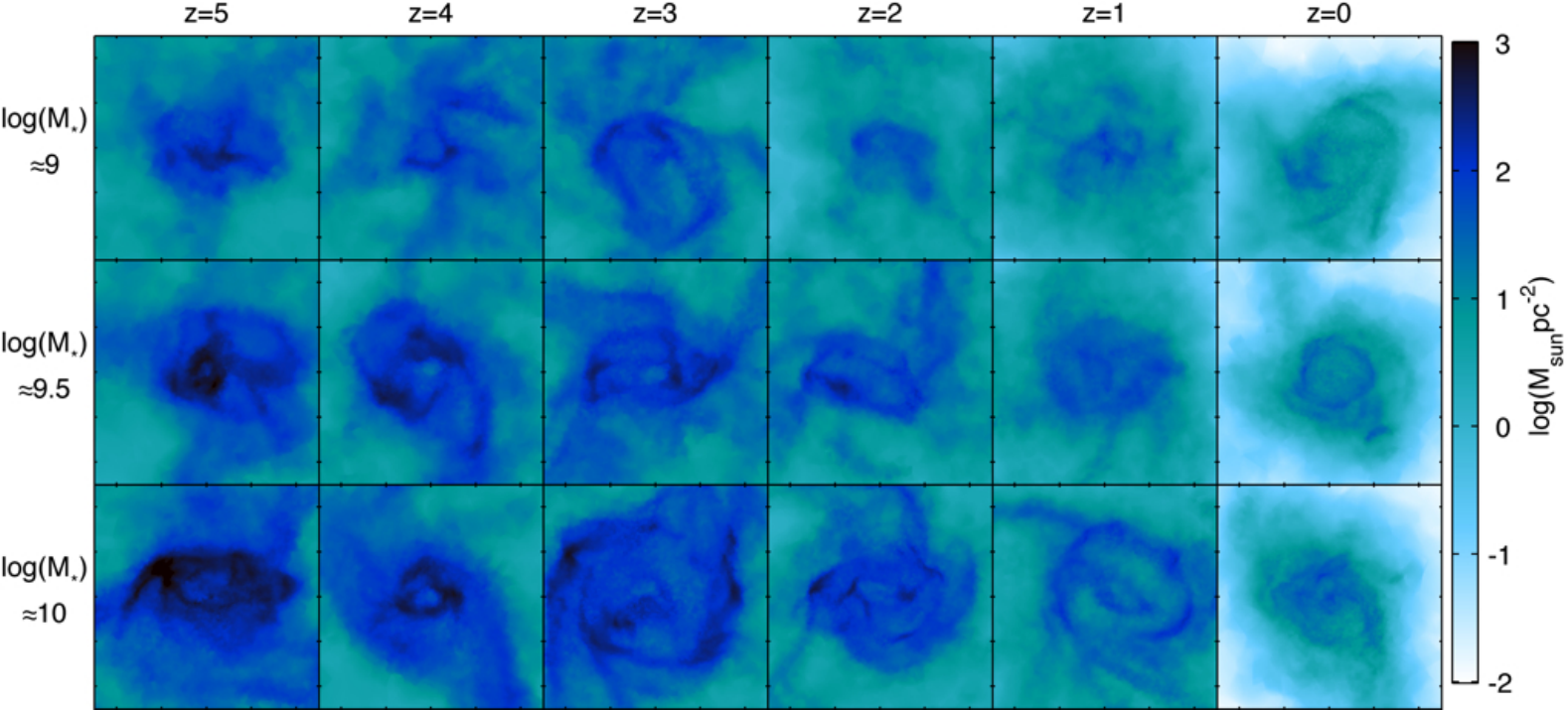}}
\subfigure[Stellar images]{
          \label{f:stellar_images_lowmass}
          \includegraphics[width=1.0\textwidth]{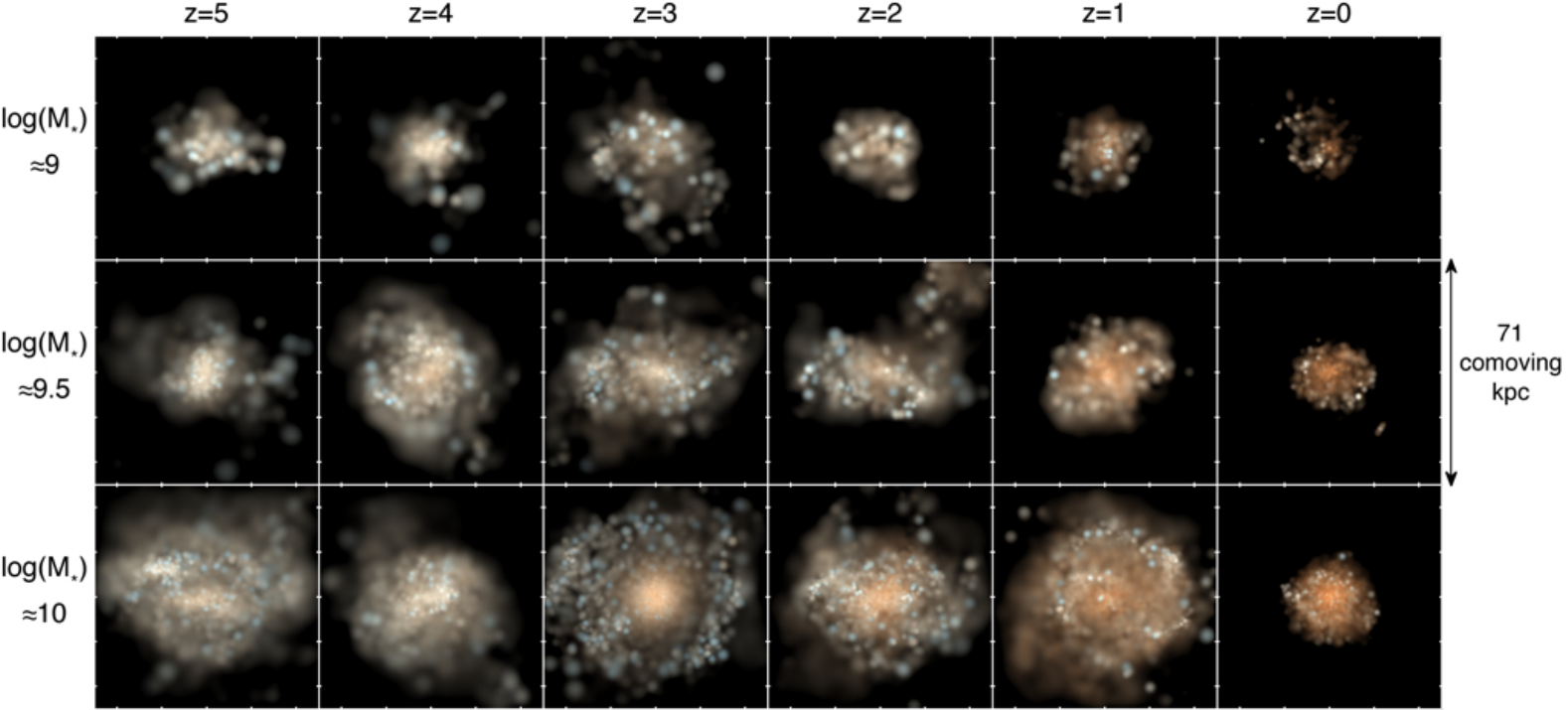}}
\caption{Galaxy images at various redshifts, from $z=5$ (left column) to $z=0$ (right column), for three low-mass bins: $M_*\approx10^{9}\Msun$ (top rows), $M_*\approx10^{9.5}\Msun$ (middle rows), $M_*\approx10^{10}\Msun$ (bottom rows). The top panel shows gas surface density in physical $\Msunpc2$, and the bottom panel shows stellar images based on the rest-frame Bgr bands. The galaxies were selected by eye to represent as much as possible the typical appearance as a function of mass and redshift. We emphasise that the galaxies in the different columns are not progenitors and descendants of each other, rather each row shows typical galaxies of a given mass as they appear at different redshifts. Each image is $71\kpc$ (comoving) on a side, and shows the galaxy face-on (based on the gas component). See \Fig{images_massive} for a similar figure for high-mass galaxies.}
\vspace{0.3cm}
\label{f:images_lowmass}
\end{figure*}

\begin{figure*}
\centering
\subfigure[Gas surface density maps]{
          \label{f:gas_images_massive}
          \includegraphics[width=1.0\textwidth]{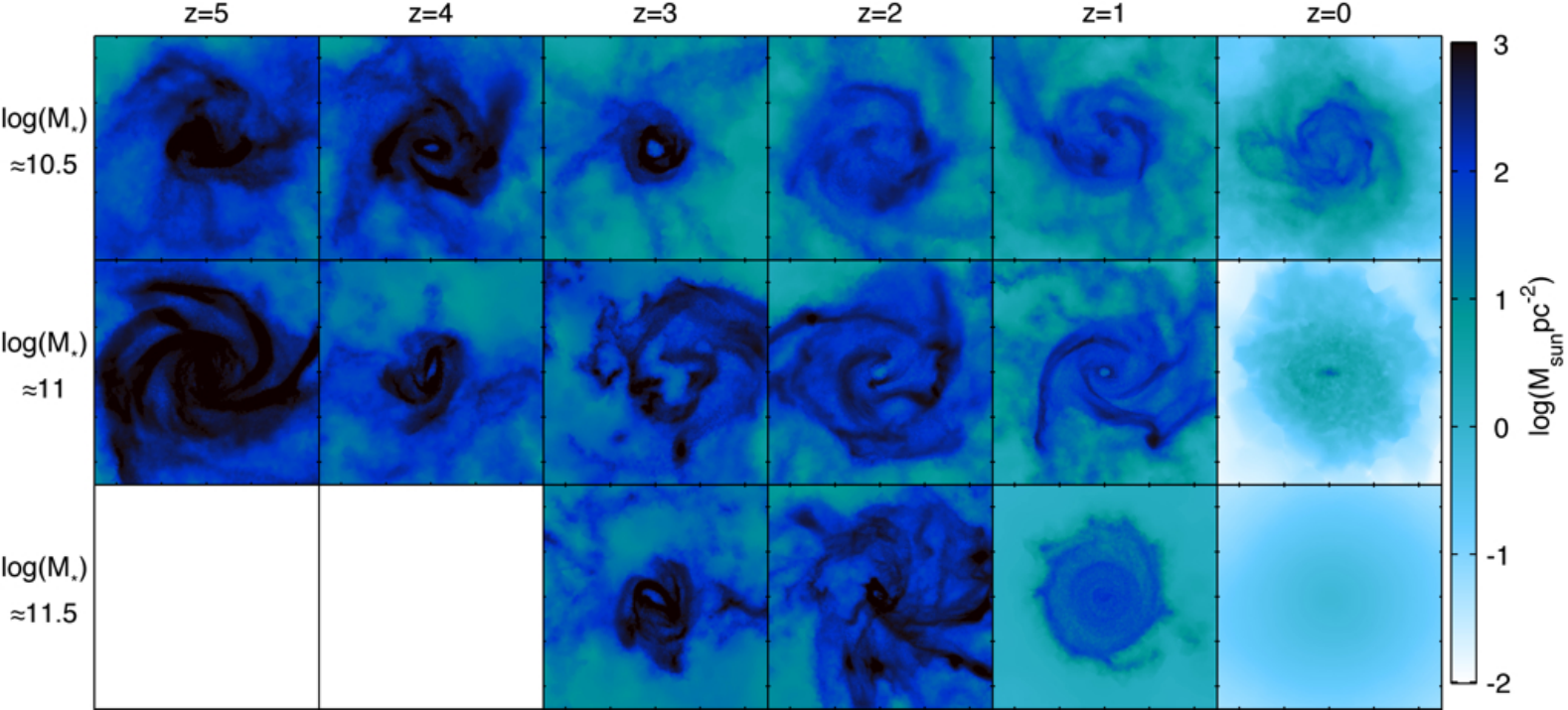}}
\subfigure[Stellar images]{
          \label{f:stellar_images_massive}
          \includegraphics[width=1.0\textwidth]{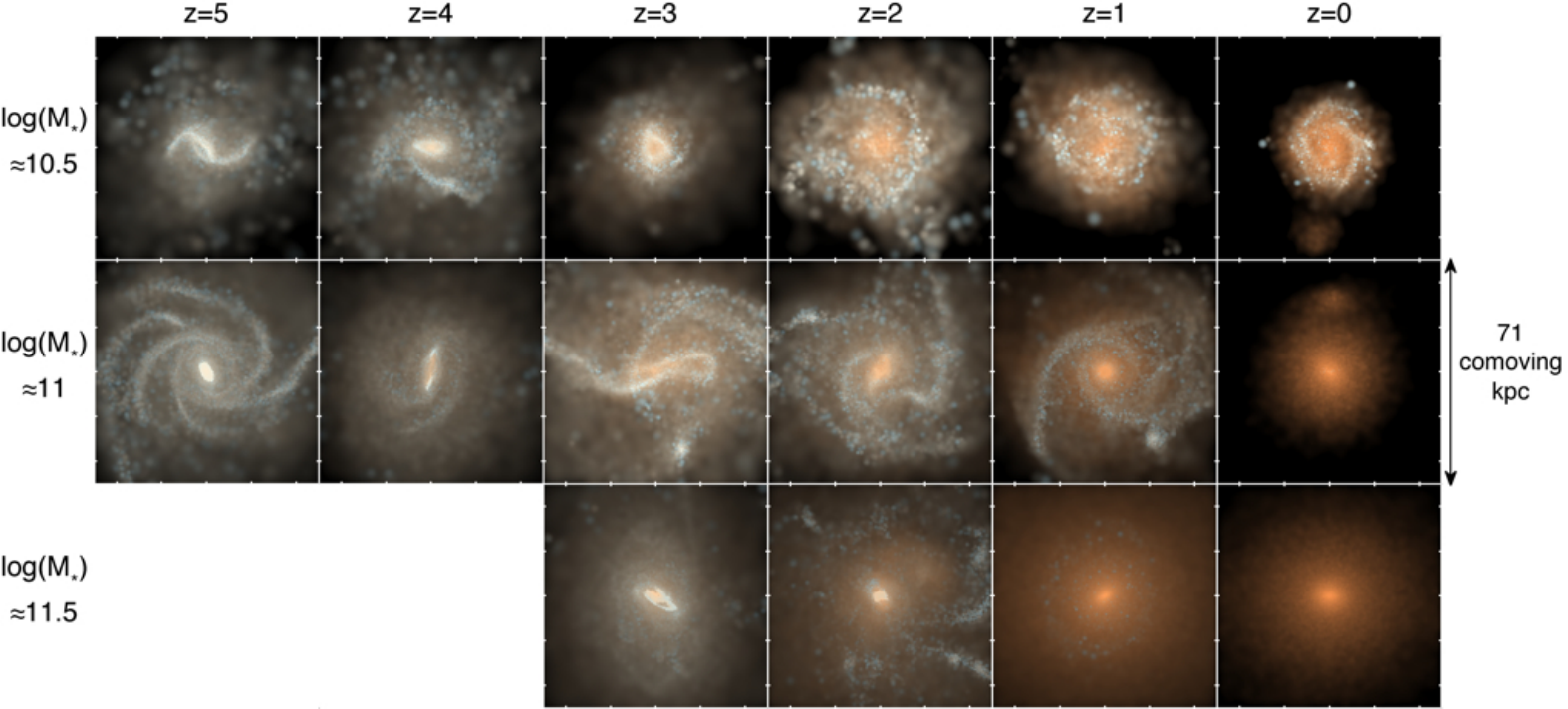}}
\caption{Same as \Fig{images_lowmass}, but for more massive galaxies: $M_*\approx10^{10.5}\Msun$ (top rows), $M_*\approx10^{11}\Msun$ (middle rows), $M_*\approx10^{11.5}\Msun$ (bottom rows). The top panel shows gas surface density in physical $\Msunpc2$, and the bottom panel shows stellar images based on the Bgr bands.}
\vspace{0.3cm}
\label{f:images_massive}
\end{figure*}

Deep observations of high-redshift galaxies reveal that galaxy morphology changes significantly with cosmic time \citep{ConseliceC_14a}. In \Figs{images_lowmass}{images_massive} we present galaxy images, both for the stellar and gas components, which show galaxies from $z=5$ to $z=0$ in a mass range of almost three orders of magnitude, from $M_*\approx10^{9-10}\Msun$ (\Fig{images_lowmass}) up to $M_*\approx10^{10.5-11.5}\Msun$ (\Fig{images_massive}). For each redshift and mass bin, we selected one `typical' galaxy, as far as possible, by visually inspecting a total of several hundreds of galaxies (a complete catalogue of stellar images for $z=0$ is presented in Torrey et al.~submitted). These images allow us to confirm that galaxy morphologies in Illustris agree, at least qualitatively, with several observed trends.

\begin{itemize}
\item
The approximate constancy of the galaxy sizes (at $z\gtrsim1$) in constant comoving units (as presented in \Figs{images_lowmass}{images_massive}) reflect a size evolution that roughly follows the cosmological scale factor $r_{\star}\propto(1+z)^{\approx-1}$. This evolution trend agrees with most observational estimates (e.g.~\citealp{LawD_12a,StringerM_14a}), however some observational studies find a weaker evolution \citep{IchikawaT_12a,StottJ_13a}. At $z=0$, the galaxies appear smaller in comoving units than at $z=1$, indicating only a slight size growth in physical units between those epochs. We will study galaxy sizes in greater detail in forthcoming work (Genel et al.~in prep.).
\item
Gas surface densities are higher towards both higher masses and higher redshifts, except at low redshift, where the trend is non-monotonic with mass, such that the most massive galaxies have in fact very low gas surface densities. These trends can be understood in terms of the mass and redshift dependencies of galaxy sizes and gas fractions, and agree well with observations. Along with the increased gas surface densities, galaxies at high redshift and in particular at high masses tend to be `clumpy', featuring $\kpc$-scale local gas density peaks, which resemble observed massive galaxy disks at $z\sim2-3$ \citep{GenzelR_11a}.
\item
Gas morphologies are only `regular', in the sense of being smooth or featuring ordered spiral arms, at low redshift $z\lesssim1$ and high masses $M_*\gtrsim10^{10.5}\Msun$. The gas distributions at all other redshift and mass bins are disturbed and asymmetric, and feature strong density contrasts. Note that these trends of increased `regularity' do not parallel the trends of the gas surface density with mass and redshift, but do parallel the dependencies of the sSFR on mass and redshift, such that galaxies with `regular' morphologies tend to have lower sSFRs.
\item
For a given stellar mass, galaxies are bluer at higher redshifts, as a result of the younger stellar populations. The trend with stellar mass is less obvious, and it is redshift-dependent. At low redshift, the most massive galaxies are clearly the reddest objects. However, low-mass galaxies are not necessarily bluer than intermediate-mass galaxies. This is related to the flattening of galaxy ages towards low masses that we show in \citet{VogelsbergerM_14a}, which disagrees with observations, and calls for attention. At $z\gtrsim2$, even high-mass galaxies are typically gas-rich and highly star-forming, and hence have rather blue colours. However, it becomes clearer towards higher masses that the blue, young stars are `painted' on top of a redder, more dominant, older stellar component. The same is true towards lower redshifts, at a given stellar mass.
\item
Even with young, blue stellar populations having a low mass-to-light ratio, which makes them come out in the images with respect to the older, redder component, the stellar images are significantly smoother and less clumpy than the gas images. We do not show here maps of stellar surface density, but it is easy to see, given the known mass-to-light ratio differences, that they would be yet smoother than the stellar light images. A quantitative study of this aspect will be presented in future work (Snyder et al.~in prep.), but it is worth noting the qualitative agreement with observations \citep{WuytsS_12a}.
\end{itemize}

\begin{figure*}
\centering
\subfigure[Face-on, No dust]{
          \label{f:HUDF_nodust}
          \includegraphics[width=0.49\textwidth]{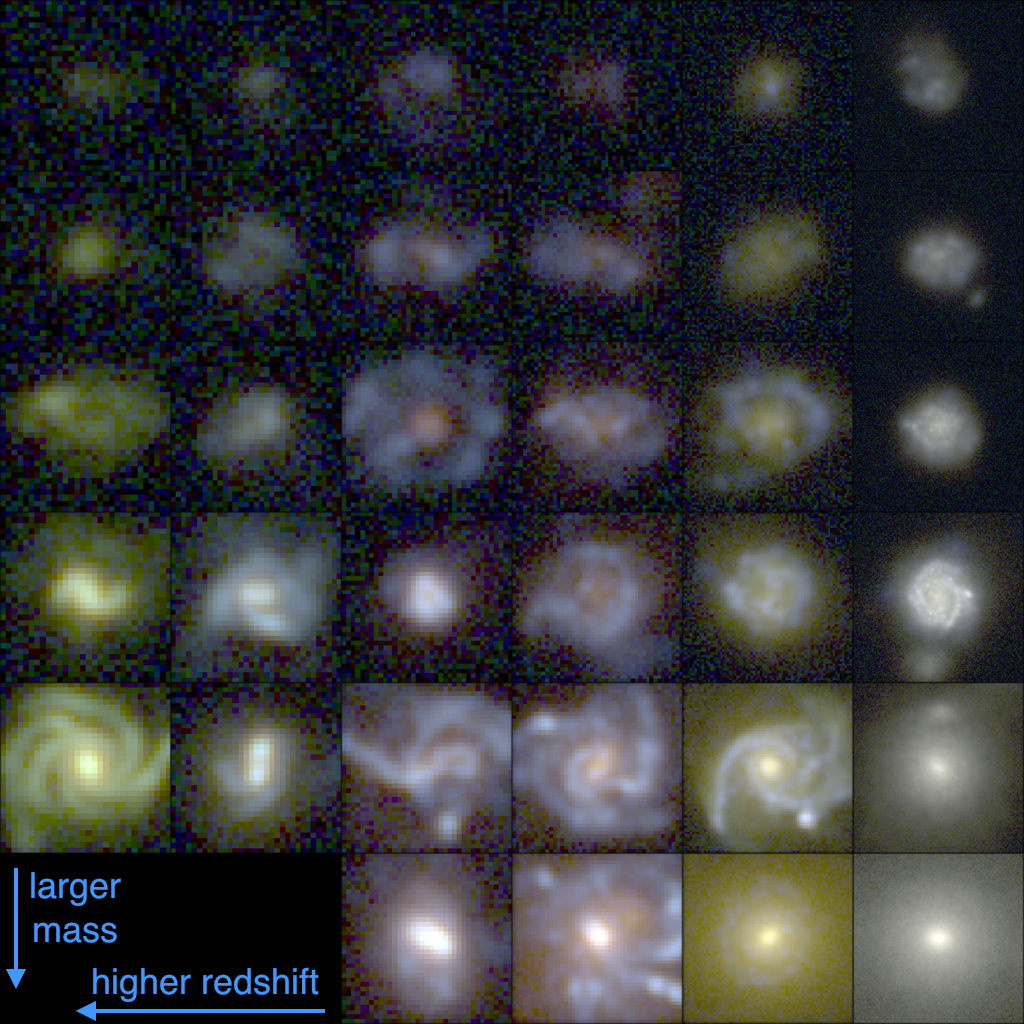}}
\subfigure[Face-on, with dust]{
          \label{f:HUDF_dust}
          \includegraphics[width=0.49\textwidth]{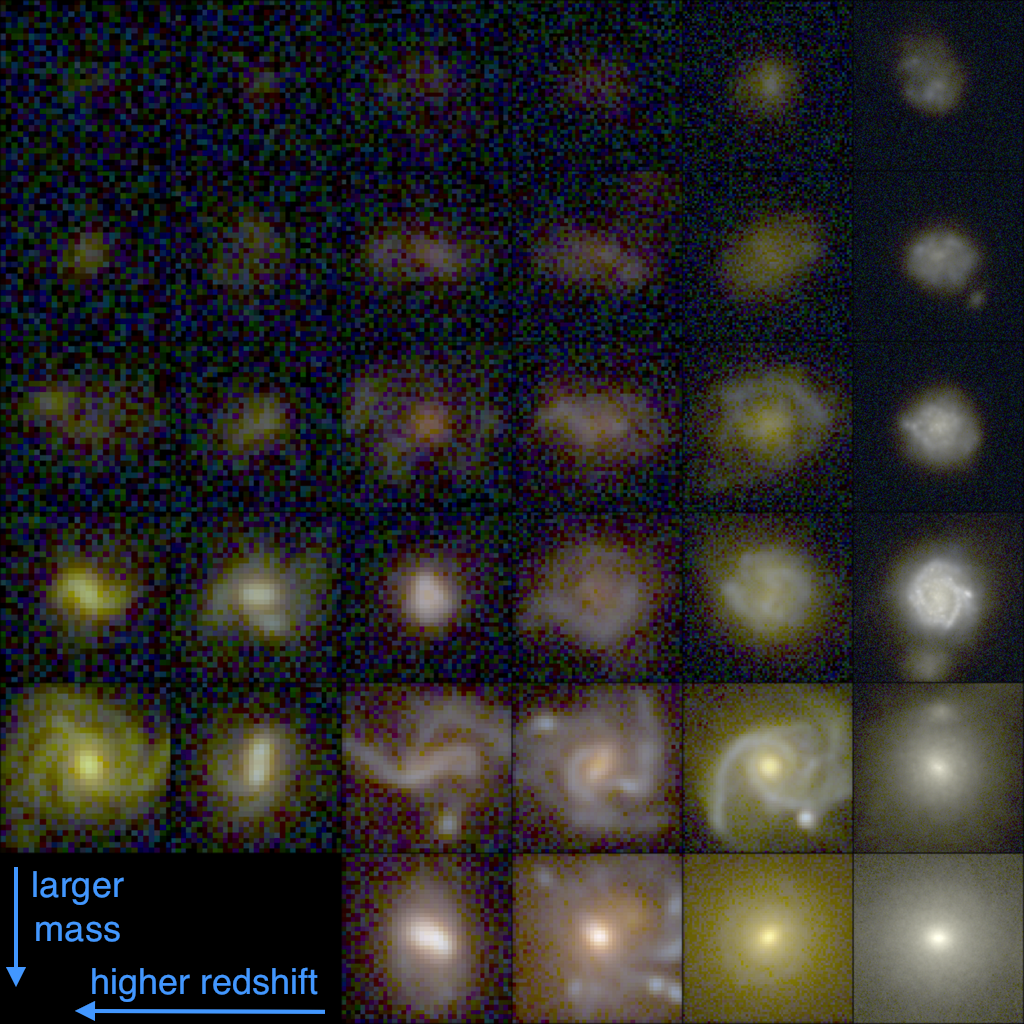}}
\caption{Mock HST observations of the galaxies presented in \Figs{images_lowmass}{images_massive}, in the observed-frame ACS/F606W (V band), WFC3/F105W (Y band), and WFC3/F160W (H band) filters, using realistic point-spread functions and noise, and shown both neglecting ({\it left}) and including ({\it right}) the effects of dust. Each image is $71\kpc$ (comoving) on a side and shows the galaxy face-on. In each panel, the columns show decreasing redshifts from left to right ($z=5,4,3,2,1,0$), and the rows show increasing stellar masses from top to bottom ($M_*=10^{9},10^{9.5},10^{10},10^{10.5},10^{11},10^{11.5}\Msun$). All panels have the same intensity stretch.}
\vspace{0.3cm}
\label{f:HUDF}
\end{figure*}

\begin{figure}
\centering
\includegraphics[width=0.475\textwidth]{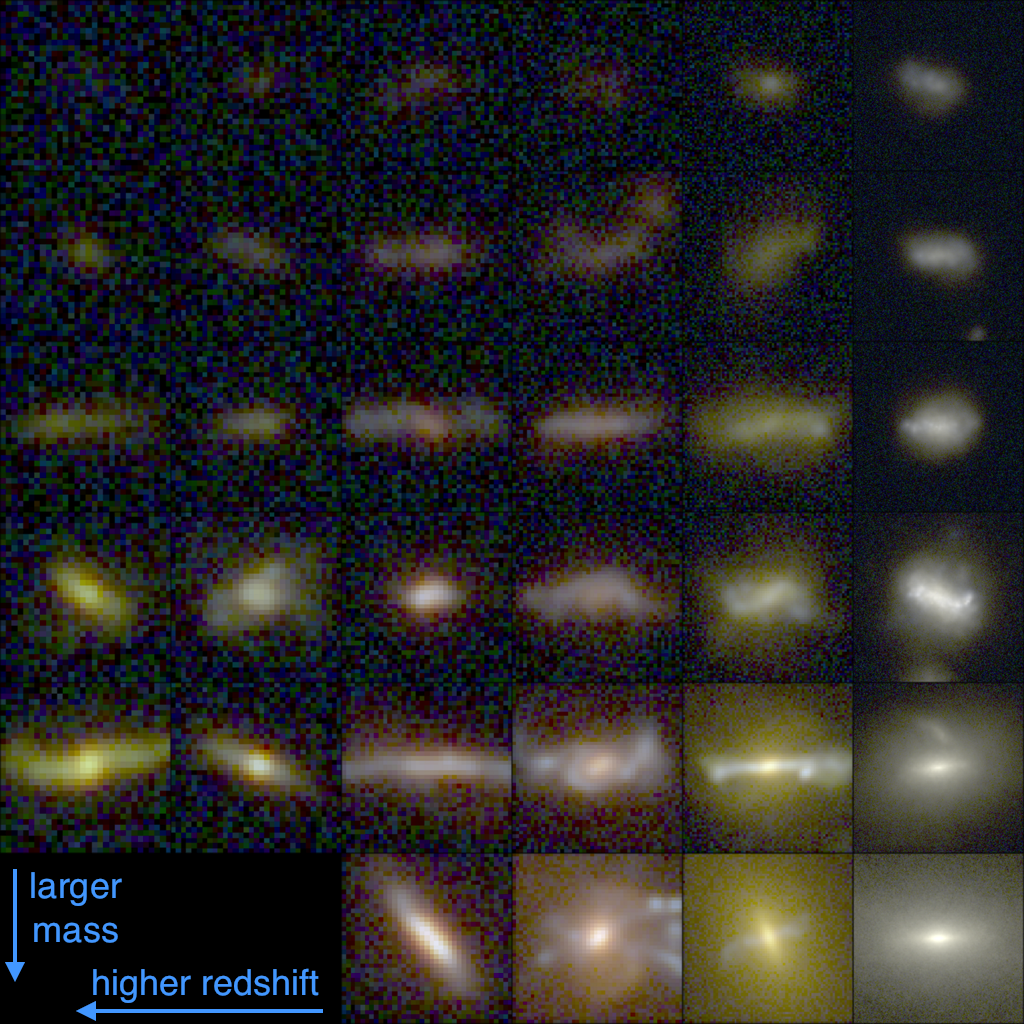}
\caption{Mock HST observations corresponding to \Fig{HUDF_dust}, only for an edge-on view.}
\vspace{0.3cm}
\label{f:HUDF_dust_edgeon}
\end{figure}

While studying galaxy morphologies as they appear in \Figs{images_lowmass}{images_massive} can reveal a lot about the evolution of galaxies, it is also important to make a direct connection with observations, and study how simulated galaxies would appear through the telescope. Creating mock observations of simulated galaxies can at the same time help to constrain the simulations more reliably and aid in the interpretation of observations of real galaxies. We will make detailed studies of this kind in future work (Torrey et al.~submitted, Snyder et al.~in prep.). Here we give a preliminary analysis, by presenting in \Fig{HUDF} mock Hubble Space Telescope (HST) broad-band observations of the galaxies shown in \Figs{images_lowmass}{images_massive}, as they would appear in the Hubble Ultra Deep Field (UDF) in constant observed-frame VYH bands. The spectral energy distribution of each galaxy is derived from its stellar populations using the \citet{BruzualG_03a} model, and then redshifted to its corresponding redshift (except for the $z=0$ galaxies, which are assumed to be at $z=0.05$). The redshift is also used to convert physical scales to angular units, which allows us then to convolve the images with Gaussian point-spread functions of $0.05$, $0.1$, and $0.2\arcsec$, for the three filters respectively, and use a pixel scale of $0.06\arcsec$, to mimic realistic HST-UDF observations (Snyder et al.~in prep.). \Fig{HUDF} also includes the effects of surface brightness dimming, Lyman continuum absorption, and the addition of Gaussian random noise with filter-dependent root mean squared values matching the total random shot noise (sky plus telescope) of an exposure similar to that of the HST XDF images \citep{IllingworthG_13a}. In \Fig{HUDF_dust} we further add the effects of dust extinction using the simple model of \citet{CharlotS_00}, which prescribes an optical depth for dust extinction as a simple function of wavelength for any given stellar population age. For young stars with age $<10^7\yr$, we use an optical depth of $\tau=1.0(\lambda/0.55\micron)^{-0.7}$, and for older stars $\tau=0.3(\lambda/0.55\micron)^{-0.7}$. This prescription was calibrated for nearby galaxies, and hence probably represents an over-estimate for the dust extinction at high redshift, and in particular at low masses, where the metallicity is significantly lower\footnote{Note however that there is also indirect evidence for extreme star-formation that is obscured at $z\gtrsim5$ \citep{StraatmanC_14a}.} \citep{BouwensR_13a,BurgarellaD_13a}. Given that and the simplicity of the model, the images in \Fig{HUDF_dust} may be considered as an order-of-magnitude exercise. \Fig{HUDF_dust_edgeon} is identical to \Fig{HUDF_dust}, except for the orientation of the galaxies, which are viewed edge-on.

The qualitative appearance of high-redshift galaxies in \Figs{HUDF}{HUDF_dust_edgeon} is similar to what is found in observations, with the exception of a dearth of very compact, unresolved, galaxies (see examples in Fig.~2 of \citet{LawD_12a}). We find diverse morphologies, which represent all the main morphological types observed in the UDF, namely disks, spheroids, clump-clusters, chain galaxies, tadpoles, mergers, and other irregular types \citep{ElmegreenD_05a,ElmegreenD_07a}. We see an increase in the fraction of disks and spheroids with cosmic time, and a corresponding decrease of irregular morphologies, as observed \citep{MortlockA_13a}. Spatial colour variations are apparent at all redshifts, more so for more massive galaxies, in agreement with observations \citep{LawD_12a,BondN_14a}. The central regions tend to be redder, at $z\sim1-4$ often surrounded by bluer `clumps', which by comparison to the gas maps can be identified as large star-forming regions, resembling observed galaxies \citep{GenzelR_08a,ElmegreenB_09a,CameronE_10a,FoersterSchreiberN_11b}.

Low-mass galaxies with $M_*\lesssim10^{10}\Msun$ at $z\gtrsim4$ are hardly resolved in these mock images, in agreement with observations \citep{ConseliceC_09a}. This is due to a combination of their smaller intrinsic sizes and dust obscuration (although it may be overestimated, see above) with the cosmological surface brightness dimming, the HST point spread function and noise. Already at $z\gtrsim2$, and in particular when dust obscuration is taken into account, a significant fraction of the surface area that is visible in \Figs{stellar_images_massive}{stellar_images_lowmass} for galaxies with $M_*\lesssim10^{10}\Msun$ becomes buried under the noise due to its low surface brightness.

It is worth commenting on two particularly interesting objects. The $z=1$ galaxy with $M_*\approx10^{11.5}\Msun$ (second from the right in the bottom row) appears in \Figs{stellar_images_massive}{HUDF} as a red galaxy with a gas disk and a subdominant disk of young stars. However, in the edge-on view in \Fig{HUDF_dust_edgeon}, the old stellar component is revealed to be disky as well, such that the system appears as two orthogonal disks with very distinct colours. This strongly suggests that the younger stellar disk formed after the main body of the galaxy assembled, probably in a `rejuvenation' phase. In addition, the appearance of the old component in the two viewing angles allows one to deduce that it is a prolate object with a rather large axis ratio of approximately $1:3$. Peculiar objects like this one can be found in Illustris thanks to the large simulated volume.

The second object worth commenting on is the most massive galaxy at $z=5$ (left column, bottom row). This is the only galaxy in the Illustris volume that has $M_*\approx10^{11}\Msun$ at $z=5$, and therefore cannot be taken as representative of any population. Nevertheless, it is a very interesting object, as it features strong spiral arms that are not regularly observed at such redshifts. While it appears extended in the images, which are $12\kpc$ on a side at $z=5$, in fact its stellar half-mass radius is $0.76\kpc$, inside which the gas surface density has an extreme value of $1.2\times10^{4}\Msunpc2$ (hence is strongly saturated in \Fig{gas_images_massive}), and the star-formation rate is $250\Msunyr$. Thus, the extended distribution that appears in the images hosts an extremely compact starburst in its center. The mean gas surface density in the region shown in the images is $2\times10^{3}\Msunpc2$, which is still very high, and under which conditions, we cannot assign high credibility to the simple dust model we use to generate \Figs{HUDF}{HUDF_dust_edgeon}. Therefore, it remains a somewhat open question what an object of this kind would look like in real observations, which calls for a more quantitative study of the possibility that current observations miss extended light distributions from such high-redshift galaxies with very large gas surface densities.

\section{Connecting galaxy populations across epochs}
\label{s:mergertrees}
\subsection{Evolution of stellar mass and merger history from $z=2$ to $z=0$}
\label{s:evolution_mass_mergers}
Cosmological simulations offer the powerful ability to follow structures, and galaxies in particular, in the time domain as they grow, and directly connect them across cosmic epochs using merger trees. This is information that can only be extracted from observations in a statistical sense, for example by using a constant number density, but even that is a difficult task whose reliability is not guaranteed \citep{BehrooziP_13c,LejaJ_13a,ToftS_14a}. Previous theoretical studies performed such exercises in a variety of interesting contexts. 

For example, \citet{SpringelV_05a} and \citet{FanidakisN_13a} investigated the question of where the descendants of bright $z\approx6$ quasars reside at $z=0$ using semi-analytical models. Other work based on semi-analytical models looked for the descendants of $z\approx2-4$ galaxy populations such as sub-millimeter galaxies, Lyman-break galaxies, and others \citep{GuoQ_09a,GonzalezJ_11a,GonzalezJ_12a}. Similar galaxy populations were studied following merger trees of DM-only simulations in combination with simple assumptions on how halos are populated by galaxies \citep{AdelbergerK_05a,ConroyC_08a,GenelS_08a}. Other studies went in the opposite direction and asked what were the high-redshift progenitors of certain selections of local galaxies \citep{FeldmannR_09a,DayalP_13a}. Thanks to the relatively large volume of Illustris, and to the fact that it was evolved down to $z=0$, we can attack such questions in a way that was not previously possible using hydrodynamical simulations (although there exist early attempts with significantly lower resolution and statistical power, e.g.~\citealp{NagamineK_02a}).

\begin{figure*}
\centering
\subfigure[Evolution of mass]{
          \label{f:progdesc_mass}
          \includegraphics[width=0.49\textwidth]{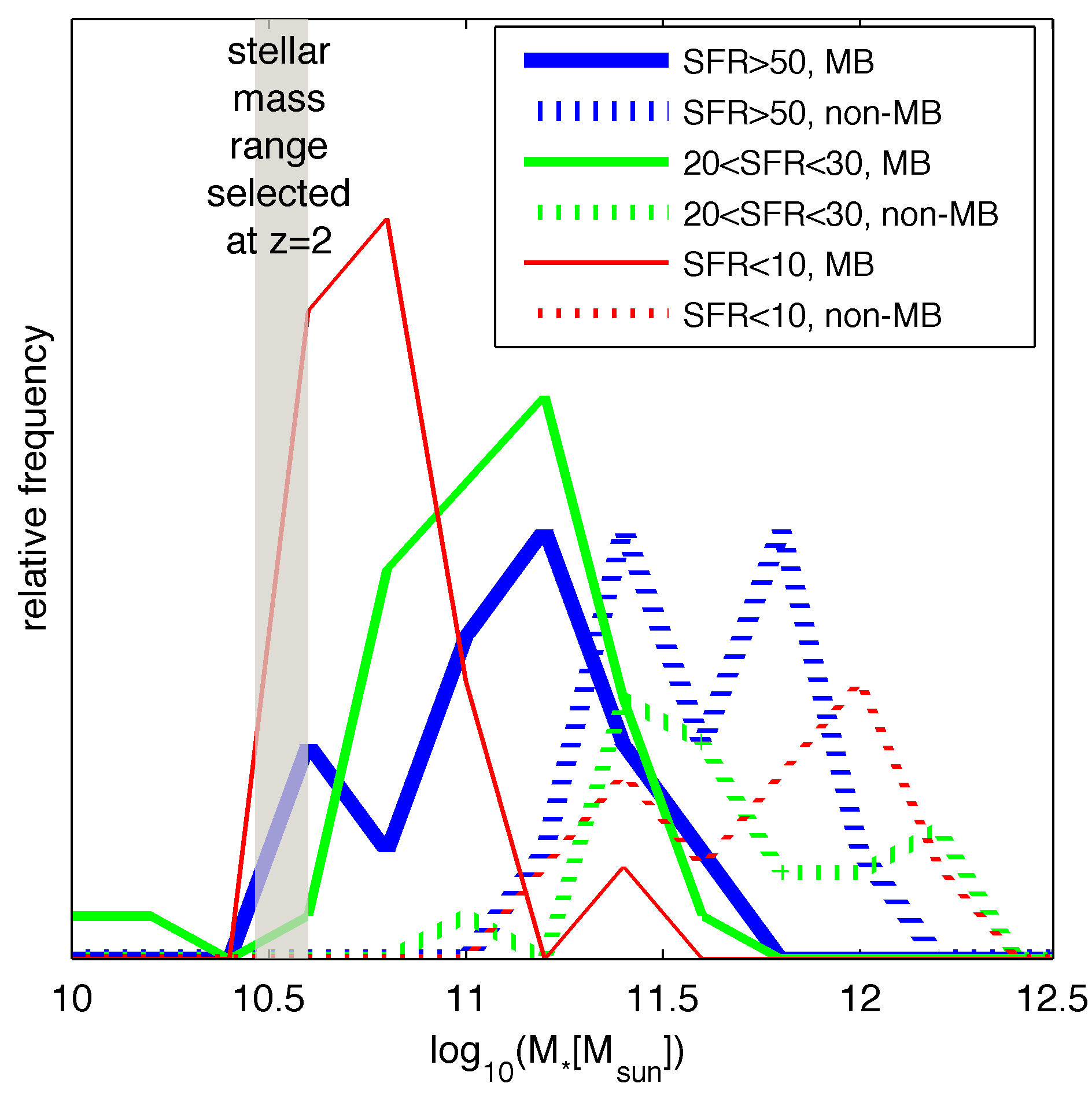}}
\subfigure[Future merging history]{
          \label{f:progdesc_nMM}
          \includegraphics[width=0.49\textwidth]{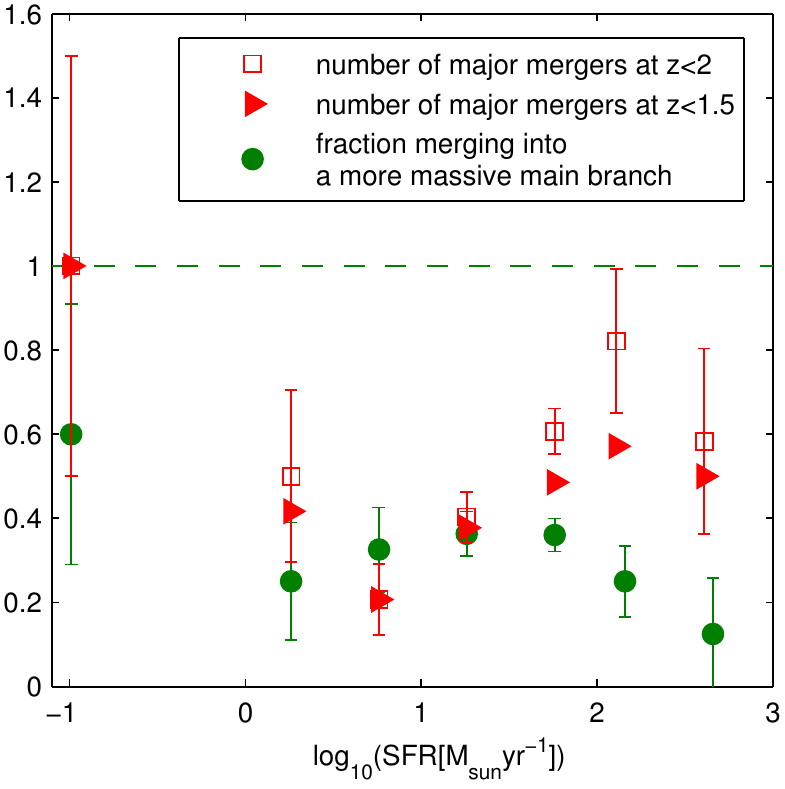}}
\caption{{\it Left panel:} The stellar mass distribution of the $z=0$ descendants of $M_*=(3-4)\times10^{10}\Msun$ galaxies selected at $z=2$, divided into three groups based on their $z=2$ star-formation rate. Between a third and a half of those $z=2$ galaxies are not on the main branch of their $z=0$ descendant (`non-MB'), i.e.~they have merged into a main branch more massive than themselves between $z=2$ and $z=0$ (dotted curves). However, the $z=0$ descendant mass of the $z=2$ galaxies that do not merge into more massive systems (solid curves) does depend strongly on the $z=2$ star-formation rate, with the `starbursts' growing significantly more than the `quiescent' ones. {\it Right panel:} Merger statistics as a function of star-formation rate for all $z=2$ galaxies with $M_*>3\times10^{10}\Msun$. The dependence of the fraction of `non-MB' galaxies is weak and not clearly statistically significant (green circles). However, for galaxies that remain on the main branch of their descendants, the mean number of major mergers between $z=2$ and $z=0$ (red squares), as well as between $z=1.5$ and $z=0$ (red triangles), depends significantly (and non-monotonically) on their $z=2$ star-formation rate. All error bars assume Poisson statistics (they are similar for the red triangles as for the squares, but not shown for visual clarity).}
\vspace{0.3cm}
\label{f:evolution_mass_mergers}
\end{figure*}

Here we use galaxy merger trees constructed from Illustris (Rodriguez-Gomez et al.~in prep.) to follow massive galaxies selected at $z=2$ to their $z=0$ descendants, and quantify their final masses and merger histories. In the first part of this exercise, we select $z=2$ galaxies in a narrow mass range $10^{10.5}\Msun<M_*<10^{10.6}\Msun$, and divide them into three populations: `quiescent' with $\SFR<10\Msunyr$, `main-sequence' with $20\Msunyr<\SFR<30\Msunyr$, and `starbursts' with $\SFR>50\Msunyr$. In \Fig{progdesc_mass} we show the stellar mass distributions of their $z=0$ descendants, where we further divide each group selected at $z=2$ into those which are on the main progenitor branch of their $z=0$ descendants (solid lines), which constitute $\approx50-70\%$ of the sample, and those which have merged between $z=2$ and $z=0$ into another galaxy more massive than themselves (dotted lines; `non-main branch'). The $z=0$ descendant mass of the galaxies that do not merge into more massive systems is found to depend strongly on the $z=2$ star-formation rate, such that the median stellar mass growth factor of `quiescent' $z=2$ galaxies is $1.75$, while for the `starbursts' it is $3.7$. Galaxies that merge into more massive galaxies before $z=0$ end up having a $z=0$ mass distribution that is consistent with being independent of their $z=2$ star-formation rate. Examining the absolute values of the $z=0$ descendant masses, we find that $\approx80\%$ of the galaxies in the `starburst' group (including both `main branch' and `non-main branch') end up at $z=0$ in galaxies with $M_*>10^{11}\Msun$, while galaxies in the `quiescent' group have a $\approx50\%$ chance of having a $z=0$ descendant with $M_*<10^{11}\Msun$.

The difference in the mass growth factors between the `starburst' and `quiescent' groups cannot be accounted for in a short time even by the high star-formation rates of the `starburst' group, since the typical sSFR at that time is $\approx1\Gyr^{-1}$. This means that either the large difference between the star-formation rates of those groups is sustained for time scales $\gtrsim1\Gyr$, or that galaxies in the `starburst' group are accreting a significantly larger mass through mergers between $z=2$ and $z=0$. In \Fig{progdesc_nMM} we show the mean number of future major mergers that galaxies undergo as a function of their $z=2$ star-formation rate (only for those galaxies that belong to the main branch of their descendants). It becomes clear that galaxies in the `starburst' group are significantly more likely to undergo a future major merger than the galaxies in the `quiescent' group\footnote{We performed similar analyses, which are not shown here, with either the stellar mass or the sSFR on the abscissa, and found significantly weaker correlations than with star-formation rate.}. This holds true also when only considering mergers at $z<1.5$, such that a direct causal relation between the merger and the enhanced SFR is eliminated. These mergers are likely responsible for a significant part of the factor of $3.7$ mass growth of the `starburst' group during this time period. This suggests that they may be residing in denser environments. It also suggests that the distribution of instantaneous $z=2$ star-formation rates is not dominated by strong `random' temporal fluctuations, but instead that those star-formation rates are tied to the growth and merger history of galaxies on long time scales (see also Sparre et al.~in prep.). Finally, the fact that galaxies with the most extreme star-formation rates (whether high or low) undergo on average $\approx1$ major merger between $z=2$ and $z=0$, but that number drops to only $<0.5$ for the more typical star-formation rates of $\approx10\Msunyr$ may suggest a close physical connection between star-formation rate enhancements and quenching, and in particular a connection between these processes and merging (e.g.~\citealp{HopkinsP_06a,HopkinsP_06b,HopkinsP_07a,HopkinsP_07b}).

\subsection{Evolution of compactness and specific star-formation rate from $z=3$ to $z=0$}
\label{s:evolution_size_sSFR}
As discussed in Section \ref{s:morphologies}, galaxy populations display a diverse morphological mix that changes with cosmic epoch. It is well established that morphological properties, such as size, concentration, and clumpiness, correlate with properties related to star-formation activity, such as sSFR, gas fraction and colour \citep{StratevaI_01a,WuytsS_11a,BellE_12a}. Such correlations are found at all redshifts where data is available, $z\lesssim3$, however they are complicated, include significant scatter, and not yet satisfactorily understood \citep{MignoliM_09a,MendezA_11a,BruceV_12a,TaliaM_13a,KavirajS_13a}. One of the basic goals of galaxy formation theory is to understand how galaxies evolve in time with respect to these parameters, and identifying the physical processes that are responsible for such transformations. In this section we explore the joint evolution of star-formation activity and size, for individual galaxies, as well as for the population of massive galaxies as a whole, from $z=3$ to $z=0$.

\begin{figure*}
\centering
\includegraphics[width=1.0\textwidth]{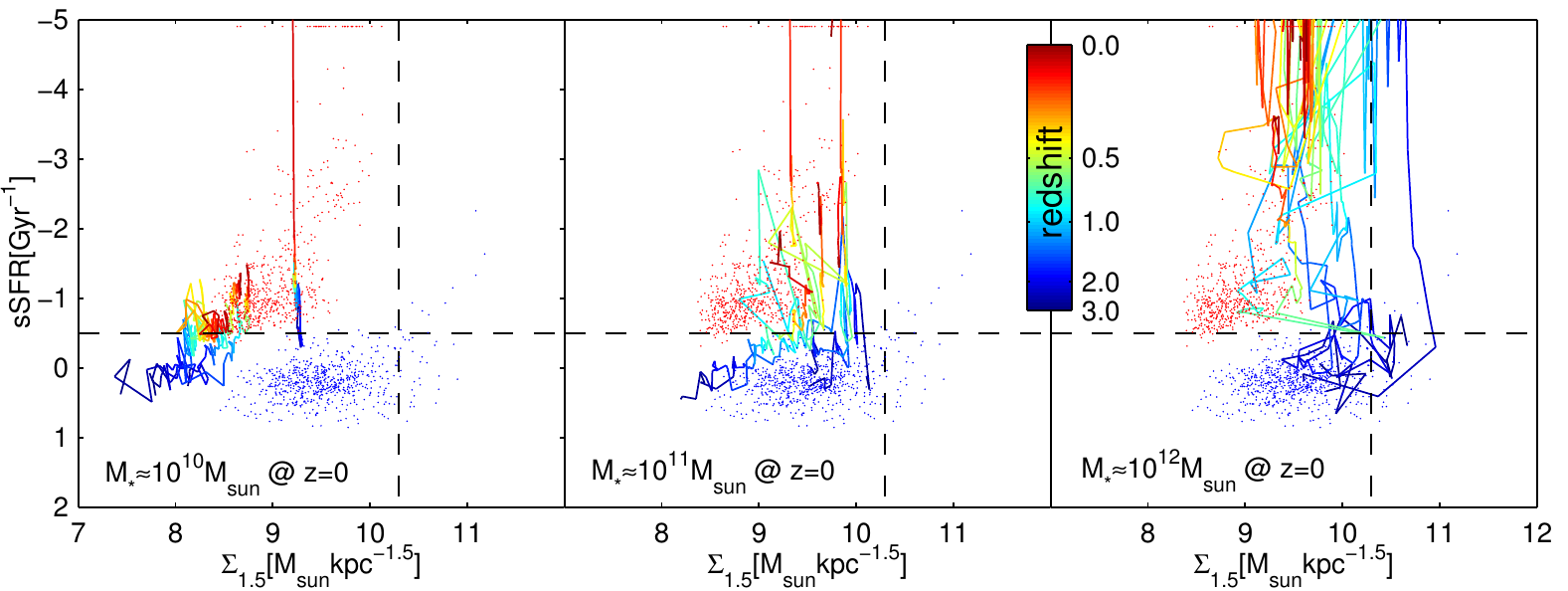}
\caption{Evolutionary tracks of individual galaxies in the plane of specific star-formation rate versus compactness ($\Sigma_{1.5}\equiv M_*/r_{\star}^{1.5}$), following \citet{BarroG_13a}. Each panel shows the tracks for $10$ galaxies selected randomly at $z=0$ in three mass bins, $10^{10}\Msun$ (left), $10^{11}\Msun$ (middle), and $10^{12}\Msun$ (right), and follows them along their main progenitor branch up to $z=3$ (red to blue). The time resolution of the tracks corresponds to the time spacing between simulation snapshots, which is approximately $1\%$ of the Hubble time, i.e.~on the order of one galaxy dynamical time. It is apparent that low-mass and high-mass galaxies evolve very differently in density and star-formation activity. In addition, two sets of identical points are overplotted in all panels, which represent a random selection of a large number of galaxies with $M_*>10^{10}\Msun$ at $z=3$ (blue) and at $z=0$ (red). The horizontal and vertical dashed lines mark the divisions between galaxies that are star-forming (bottom) and quiescent (top), as well as between extended (left) and compact (right) galaxies, as defined by \citet{BarroG_13a}. Galaxies with ${\rm sSFR} < 10^{-4.9}\Gyr^{-1}$ are assigned that value such that they do not fall outside the limits of the plot (red dots).}
\vspace{0.3cm}
\label{f:size_sSFR_evolution}
\end{figure*}

\Fig{size_sSFR_evolution} shows the plane of sSFR versus `compactness' defined as $\Sigma_{1.5}\equiv M_*/r_{\star}^{1.5}$. This parameter space is split, following \citet{BarroG_14a}, into four quartiles (delineated by the dashed lines): `quiescent' (top) versus `star-forming' (bottom), and `extended' (left) versus `compact' (right). Galaxies in Illustris with $M_*>10^{10}\Msun$ appear as small dots in all panels, for $z=3$ (blue) and for $z=0$ (red). Evidently, galaxies at different times populate distinct loci on this plane. Galaxies at $z=0$ are generally less star-forming and more extended than galaxies at $z=3$. Only a small fraction of the galaxies at $z=3$ are in the `compact' region of the plane, while at $z=0$ none are. \citet{BarroG_14a} find the same qualitative evolutionary trends using HST observations at $0.5<z<3$. While we do not show this explicitly here, we note that also quantitatively, the observed galaxy number densities in the various quartiles are reproduced in Illustris with differences of less than a factor $\approx2$.

Interpreting their results, \citet{BarroG_14a} suggest that galaxies may evolve on two distinct evolutionary paths (see their Fig.~6). First, some galaxies evolve on an `early track' that involves first compactification, then rapid quenching, and finally a gradual size growth on the red sequence. Second, some galaxies evolve on a `late track' that never goes through a compact phase, and instead just gradually become more and more quiescent. \citet{DekelA_14b} provide a theoretical framework for these two scenarios, where gas-rich disks (corresponding to low spin parameters) undergo disk instabilities that make them contract and evolve on the `early track', while less gas-rich disks quench on the `late track' without contracting first. Their model, however, does not intrinsically include the required quenching mechanism(s).

In \Fig{size_sSFR_evolution} we confirm the basic interpretation that \citet{BarroG_14a} put forward for their observations, albeit with important variations to their proposed scenario, by following individual galaxies as they evolve in the simulation. To do so, we select galaxies at $z=0$ in three groups based on their stellar mass, with $10$ random galaxies in each group, and follow their main progenitor histories up to $z=3$. The evolutionary tracks of the galaxies in each mass group are shown in a separate panel in \Fig{size_sSFR_evolution}, from $10^{10}\Msun$ on the left to $10^{12}\Msun$ on the right. The colour along each track represents redshift, from the time of selection $z=0$ (red) up to $z=3$ (blue). It becomes clear that galaxies of different masses populate completely different regions of this parameter space throughout their full formation histories. In particular, massive galaxies (right panel) evolve in a path reminiscent of the `early track' hypothesised by \citet{BarroG_14a}, where they cross the `quenching' line already at $z\approx2$, and evolve non-monotonically in the horizontal direction -- first undergoing compactification, thereafter expansion. In contrast, galaxies selected at $z=0$ with $M_*\approx10^{10}\Msun$ (left panel) evolve monotonically, and far less rapidly, in accordance with the \citet{BarroG_14a} `late track'. Their compactness evolves steadily with time towards higher $\Sigma_{1.5}$ as they grow their mass, and their sSFRs drop gradually, such that they cross the `quenching' line\footnote{Note that most galaxies in this mass bin have sSFRs of $\approx0.1\Gyr^{-1}$, such that they are in fact on the star-formation main sequence for $z=0$. Thus, they are not formally `quenched' as their massive $M_*\approx10^{12}\Msun$ counterparts. However, for the purposes of this section, we follow the terminology of \citet{BarroG_14a} and define a `quenching' line at a sSFR of $0.3\Gyr^{-1}$.} as late as $z\sim0.5$.

Hence, we find that galaxies in Illustris indeed evolve along similar tracks to those envisioned by \citet{BarroG_14a} and \citet{DekelA_14b}. In addition, we identify the different tracks with galaxies of different stellar masses. Importantly, however, we find that the two tracks are not bimodal. This is demonstrated in the middle panel of \Fig{size_sSFR_evolution}, which shows that galaxies of intermediate masses evolve on an `intermediate track', where they undergo compactification from $z=3$ to $z\sim1$, which is then followed by gradual quenching at $z\lesssim1$. Future observational tests of this mass segregation will serve as an interesting test of our simulation. 

We interpret the continuous spectrum of evolutionary trends in terms of continuities in the significance of different physical processes as a function of galaxy mass. With respect to the quenching mechanisms, the dominant process for the most massive galaxies that undergo rapid, early, and strong quenching is AGN feedback \citep{DiMatteoT_05a,SpringelV_05d}. We have shown evidence for this in Section \ref{s:SMF_MSMH}, as well as in \citet{VogelsbergerM_13a} and \citet{TorreyP_14a}. At lower masses, the role that AGN feedback plays diminishes, and indeed most galaxies with $M_*\approx10^{10}\Msun$ lie on the main sequence of star-forming galaxies. There is no sharp transition, but instead a continuous trend of quenching redshift with stellar mass. With respect to the evolution on the compactness axes, we find that stellar mass growth by star-formation is accompanied by an increase in $\Sigma_{1.5}$, such that galaxies in the bottom part of the plane move to the right as they evolve. In addition, some galaxies, after being quenched, undergo expansion due to a process that we do not directly identify here, but is likely to be related to dry mergers \citep{NaabT_06a,NaabT_09a}. This becomes increasingly important at higher masses, such that the most massive galaxies evolve towards lower $\Sigma_{1.5}$ after being quenched, intermediate-mass galaxies evolve vertically in \Fig{size_sSFR_evolution}, and for the least massive galaxies we consider, this process seems to be sub-dominant with respect to growth by in-situ star-formation, such that they continue evolving towards higher $\Sigma_{1.5}$ even at late times. As for the quenching case, we find no sharp transition between two evolutionary tracks. While such sharp transitions or bimodalities are common and probably useful, in various contexts, for interpreting observational data and constructing analytical models, hydrodynamical cosmological simulations often show more subtle, smoother transitions between limiting regimes (e.g.~\citealp{NelsonD_13a}). A further study of the physical mechanisms determining the size evolution and star-formation histories of massive, compact high-redshift galaxies will be presented in Wellons et al.~(submitted).

\section{Summary and conclusions}
\label{s:summary}
In this paper we discussed several aspects of galaxy evolution across cosmic time in the Illustris Simulation, made comparisons between the simulation and observations for various epochs, and gave predictions for future observations. Illustris, along with other recent simulations \citep{HirschmannM_14a,KhandaiN_14a,SchayeJ_14a}, represents a new generation of cosmological hydrodynamical simulations, which can for the first time be compared against observations on the basis of large samples of simulated galaxies that span the entirety of cosmic history. This type of simulations can therefore be constrained by observations in a reliable, unbiased way. 

Specifically in Illustris, which is also the largest to date (\Fig{simulations_history}), the physical models have been tuned to roughly reproduce two basic observables of our Universe, namely the present-day stellar mass function and the history of cosmic SFR density. In two companion papers \citep{VogelsbergerM_14a,VogelsbergerM_14b} we show that the simulation reproduces many additional observables of the local Universe. In this paper we expand the scope, and show that also several basic properties of the high-redshift galaxy populations in Illustris are matched well against observations. This gives us confidence that the simulation can be used to learn about how galaxies form in the real Universe. We also identify tensions with observations. These will be used in future work to improve the physical modelling of the processes governing galaxy formation. Our main findings are recapped below.

\begin{itemize}
\item
{\bf Galaxy stellar masses.} We find that the stellar mass functions and relations between stellar mass and halo mass at redshifts $z=0-7$ are generally in good agreement with observations (Section \ref{s:SMF_MSMH}). This may indicate that the combination of gas cooling and accretion, in concert with gas ejection induced by feedback processes, are broadly captured in our simulation, in both the physical and numerical sense. This result is numerically robust, in the sense that it is converged with respect to resolution \citep{TorreyP_14a}, and in the sense that our $(106.5\Mpc)^3$ simulation volume is large enough such that the uncertainty due to cosmic variance is small compared to the observational uncertainties, as we show in Section \ref{s:cosmic_variance}. As the details of the physical modelling were calibrated to roughly reproduce galaxy masses in the local Universe, it is a remarkable result that the stellar mass functions at higher redshifts are in fact in an even better agreement with observations than at $z=0$. Nevertheless, several important tensions with observations do exist, in particular at low redshift. Even though the feedback processes we employ are very energetic, additional suppression of galaxy stellar masses is probably required both in low-mass and in high-mass halos. We find it very plausible that higher energetics will not solve the problems easily, and may in fact introduce new discrepancies, as discussed below. Therefore, we believe that future research is needed to develop improved algorithms for sub-resolution feedback prescriptions that will capture reality more faithfully and lead to increased efficiency of the suppression of star-formation (e.g.~\citealp{HopkinsP_14a}).
\item
{\bf Halo gas content.} We find that at low redshift, massive halos of $\sim10^{13}\Msun$ are almost devoid of gas as a result of radio-mode AGN feedback, in disagreement with observations (Section \ref{s:gas_halos}). Hence, there remains no freedom in changing the energetics alone of AGN feedback in our model to match both the stellar and the gaseous contents of massive halos (see also Sijacki et al.~(in prep.), where a detailed account of black hole properties in Illustris is given). This probably means that other aspects of our implementation of AGN feedback require modification. For example, its duty cycle may be too low, such that short and strong bursts are able to expel large amounts of mass to large distances while not suppressing residual star-formation completely. Some AGN feedback models, indeed, obtain reasonable stellar masses in massive halos while in fact having the opposite problem of not suppressing the gas mass {\it enough} \citep{MartizziD_13a,PlanellesS_13a}\footnote{It should also be kept in mind that matching stellar and gas fractions at $z=0$ for a set of `zoom-in' massive halos does not guarantee that the model reproduces correct stellar masses for lower-mass halos and at higher redshifts as well. As the baryon conversion efficiency of halos of different masses are not independent of each other, only a full cosmological box can truly constrain any particular AGN feedback model.}. It is also possible that a simpler modification, such as reducing the total energetics of AGN feedback, would result in more realistic gas properties. However, in such a case, other processes that we are currently neglecting will have to be modeled in order to suppress the buildup of stellar mass. Such processes may include non-equilibrium metal line cooling in AGN proximity zones \citep{OppenheimerB_13b}, cosmic ray pressure support \citep{SijackiD_08a}, and thermal conduction \citep{RuszkowskiM_11a}.
\item
{\bf Galaxies around massive hosts.} The stellar content of satellite galaxies in massive halos is found to be in good agreement with observations (Section \ref{s:z0MassiveEnd}). The masses and spatial distribution of satellites inside and around massive halos can serve as strong tests of theoretical models, as they are sensitive to both gravity, hydrodynamics, and feedback processes. Hence, we find it to be an important success of our simulation that it reproduces the observed profiles of galaxies around massive hosts (Section \ref{s:satellites}). We find that in physical units, and when satellite are selected using a constant mass ratio, the evolution of the galaxy number density profiles is weak, in agreement with existing observations up to $z\approx1.6$. We find that this non-evolution extends to even higher redshifts $z\gtrsim3$, and give quantitative predictions in this regard. We also give predictions for a different selection of satellite galaxies, based on a constant magnitude, which extend existing observations to a larger redshift range.
\item
{\bf Circular velocity profiles.} We study galaxy circular velocity profiles as a function of galaxy stellar mass, redshift, and radius, out to $R_{200c}$ (Section \ref{s:gas_halos}). We find that $z=0$ galaxies in Illustris have rising or flat rotation curves inside the radius where baryons dominate the mass, typically a few percent of $R_{200c}$. The peak value of the circular velocity is however larger than $V_{200c}$, as a result of baryon condensation, and in agreement with observations. For the most massive galaxies at $z\gtrsim1$, however, we predict that the stellar component is more concentrated, which drives a peak in the circular velocity profile at smaller radii, around $0.01R_{200c}$. We also find that the dark matter halo responds to baryonic effects in a non-monotonic mass-dependent way, which reaches peak contraction for galaxies with $M_*\approx10^{11}\Msun$ at $z=0$ and for higher stellar masses at higher redshifts. Due to our implementation of a `smooth' star-formation feedback, low-mass galaxies do not alter the structure of their host halos, and in particular do not create a dark matter core.
\item
{\bf Galaxy gas fractions.} We study galaxy molecular gas fractions as a function of stellar mass and redshift in Section \ref{s:gas_galaxies}. We find good agreement with observations within the various modelling and observational uncertainties, up to the highest measured redshifts, at $0<z\lesssim3$. We find increasing gas fractions with redshift up to $z\approx2.5$, above which the increase stops and the dependence on redshift flattens, while the mass dependence steepens. These predictions are testable against upcoming observations.
\item
{\bf Specific star-formation rates.} We quantify the evolution of galaxy specific star-formation rates (Section \ref{s:sSFR}), and find that they are in generally good agreement with observations all the way up to $z\approx8$. However, the tension that is common to all hydrodynamical simulations, and almost all theoretical models in general, namely that the specific star-formation rates at $z\sim1-3$ are underestimated, remains in Illustris as well. We find that the specific star-formation rates in Illustris follow very closely the specific accretion rates of dark matter halos, albeit with a normalization offset of a factor $\approx3$ towards higher specific star-formation rates. Understanding this curious similarity will be the subject of future work.
\item
{\bf Galaxy morphologies.} We present a qualitative account of galaxy morphologies from $z=0$ to $z=5$ by presenting images of typical galaxies as a function of mass and redshift. In addition, we create mock images of these selected galaxies as they would realistically appear in the Hubble Ultra Deep Field (Section \ref{s:morphologies}). We find, in agreement with observations, that galaxies become more irregular and clumpy towards higher redshift, with increasing gas surface densities. We also identify the buildup of red bulges towards higher masses and lower redshifts.
\item
{\bf Time evolution of individual galaxies in mass, size, and star-formation rate.} We follow galaxy populations in time using merger trees, and study their evolutionary tracks (Section \ref{s:mergertrees}). In one application, we select galaxies at $z=2$ and study their mass evolution and major merger histories down to $z=0$. We find that $z=2$ galaxies that are above the star-formation main sequence are far more likely to undergo a major merger by $z=0$ than galaxies that are on the main sequence itself at $z=2$, and that their $z=0$ descendants are typically more massive. Interestingly, also the most quenched $z=2$ galaxies (those far below the star-formation main sequence) have higher chances of undergoing a future major merger. These trends suggest a similarity between extreme star-formation rates and quenching, that both are related to the large-scale environments of galaxies and to their long-term formation histories. We also study the evolution of galaxies in the parameter space of their star-formation activity and their compactness, and find a strong mass segregation in the nature of size evolution and quenching time scales. The Illustris simulation illuminates the distinct physical processes at play, and provides a powerful framework for interpreting current and future observations of galaxies across cosmic time.
\end{itemize}

\section*{Acknowledgements}
We are grateful to Andrei Kravtsov and Tomer Tal for providing us with their data and their helpful interpretations. We would like to thank Guillermo Barro, Peter Behroozi, Nicolas Bouch{\'e}, Reinhard Genzel, Chung-Pei Ma, Federico Marinacci, Pascal Oesch, and Linda Tacconi for useful discussions. We acknowledge comments on an earlier draft from Stefano Andreon, Kyoungsoo Lee, and Ali Rahmati. Simulations were run on the Ranger and Stampede supercomputers at the Texas Advanced Computing Center as part of XSEDE project TG-AST110016, the CURIE supercomputer at CEA/France as part of PRACE project RA0844, and the SuperMUC computer at the Leibniz Computing Centre, Germany, as part of project pr85je. VS acknowledges support by the DFG Research Centre SFB-881 `The Milky Way System' through project A1, and by the European Research Council under ERC-StG grant EXAGAL-308037. GS acknowledges support from the HST grants program, number HST-AR-12856.01-A. Support for program \#12856 (PI J.~Lotz) was provided by NASA through a grant from the Space Telescope Science Institute, which is operated by the Association of Universities for Research in Astronomy, Inc., under NASA contract NAS 5-26555. LH acknowledges support from NASA grant NNX12AC67G and NSF grant AST-1312095.

\label{lastpage}

\end{document}